\documentclass[reprint, amsmath, amssymb,onecolumn, pra]{revtex4-1} 

\usepackage[bookmarks=false,linkcolor=blue,urlcolor=blue,colorlinks,citecolor=blue]{hyperref}

\pdfoutput=1 
\usepackage{graphicx}  
\usepackage{dcolumn}   
\usepackage{bm}        
\usepackage{amssymb}   

\newcommand{\be}{\begin{equation}}
\newcommand{\ee}{\end{equation}}
\newcommand{\bea}{\begin{eqnarray}}
\newcommand{\eea}{\end{eqnarray}}

\usepackage{xcolor}

\usepackage[english]{babel}
\usepackage{amsmath,amsthm,amssymb}
\usepackage{amsfonts}
\usepackage{amsopn}
\usepackage{float}

\usepackage{url}

\usepackage{graphicx}
\usepackage{dcolumn}
\usepackage{bm}
\usepackage{color}

\usepackage{float}

\newcommand{\bpm}{\begin{pmatrix}}
\newcommand{\epm}{\end{pmatrix}}
\newcommand{\bmm}{\begin{matrix}}
\newcommand{\emm}{\end{matrix}}

\begin{document}

\author{Damien Barbier}
\affiliation{Sorbonne  Universit\'e,  Laboratoire  de  Physique  Th\'eorique  et  Hautes  Energies, CNRS, UMR 7589,  4  place  Jussieu,  Couloir  13-14,  5\`eme   \'etage,  75252 Paris Cedex 05,  France}

\author{Leticia F. Cugliandolo}
\affiliation{Sorbonne  Universit\'e,  Laboratoire  de  Physique  Th\'eorique  et  Hautes  Energies, CNRS, UMR 7589,  4  place  Jussieu,  Couloir  13-14,  5\`eme   \'etage,  75252 Paris Cedex 05,  France}
\affiliation{Institut Universitaire de France,    1 rue Descartes, 75231 Paris Cedex 05,  France}

\title{A constrained TAP approach for disordered spin models:\newline
application to the mixed spherical case}

\begin{abstract}
We revisit the metastability properties of the mixed $p$-spin spherical disordered models. Firstly, using known methods, we show that there is temperature chaos in a broad range of temperatures.
\newline
Secondly, we modify the definition of the Thouless-Anderson-Palmer free energy density by including constraints that enforce a chosen overlap between the searched metastable states and another reference state, that could be a characteristic one of a different temperature. We argue that this refined analysis provides clues to understand the weird behaviour of the low temperature relaxation dynamics of these models, and suggests ways to improve the treatment of the initial conditions to overcome the difficulties encountered so far.
\end{abstract}


\maketitle

\tableofcontents

\section{Introduction}

The pure (monomial) $p$-spin disordered  spherical model is a  solvable classical system that has been the focus of intense study since it appeared in the literature in the early 90s. Its  static~\cite{Crisanti1992}, metastable~\cite{Cavagna1998,Kurchan1993} and 
dynamic~\cite{Cugliandolo1993} properties can be obtained,
in the thermodynamic limit, with analytic methods (namely, the replica trick, the Thouless-Anderson-Thouless approach
and the Schwinger-Dyson equations coupling linear response and correlation functions). A rather complete and consistent picture emerges from these studies.  
In particular,  the model realises the random first order phase transition scenario
and, for this reason, it is accepted as the simplest model for fragile glass physics~\cite{Berthier2011,Bouchaud1996,Cavagna2009,Cugliandolo2002}. 

An easy but intriguing generalisation consists in adding two different pure spherical models, with potential energies involving interactions between different number of spins and, for concreteness, both strictly larger than two. This construction yields a mixed $p$-spin, still spherical, disordered model. One reason for being interested in these generalisations is that, in the glassy context, they extend the mode-coupling approach developed to describe the dynamics above the dynamic critical temperature $T_d$ and capture richer relaxations of the correlation functions~\cite{Berthier2011,Bouchaud1996,Cavagna2009,Cugliandolo2002}. Another reason is that, in mappings between optimisation problems and disordered spin systems, models with several $p$-spin terms naturally arise~\cite{Monasson1997}. Finally, one can simply be interested in the behaviour of such an extended Hamiltonian.

Standard knowledge on the metastability of the spherical $p$-spin models suggests that the behaviour of the mixed case should be different from the one of the monomial model in many respects. Indeed, a simple and very convenient property of the equilibrium and metastable states of the pure model is lost. In the monomial model, due to the homogeneity of the Hamiltonian, the states can be followed in temperature until the spinodal at which they disappear without crossing, merging nor dividing~\cite{Kurchan1993}. In other words, there is no chaos in temperature. In particular, the states that dominate the equilibrium properties are the same in the whole low temperature phase~\cite{Kurchan1993,Barrat,Barrat1997}. This simple structure is lost in the mixed case. 

The static properties of the mixed model, derived with the replica method, remain very similar to the ones of the pure model: at a critical temperature $T_s$ the replica symmetry is broken into a one-step replica symmetry breaking form signaling the equilibrium transition from the disordered paramagnetic phase to the low-temperature glassy one~\cite{Crisanti2006}. The relaxation dynamics from totally random initial conditions, mimicking equilibrium at infinite temperature, indicate the existence of a dynamic transition at a higher temperature $T_d$, below which the evolution is forced to remain out of equilibrium in the infinite system size limit~\cite{Cugliandolo1996}, taking place on a threshold level, at higher (free) energy density than the equilibrium one.

However, in line with chaotic structures, the early works on the mixed model~\cite{Barrat1997} already showed peculiar behaviour. In particular, the dynamics of initial states in equilibrium at $T'\in [T_s, T_d]$ quenched at very low temperatures $T<T_{\rm RSB}(T')$ showed ageing phenomena at energy levels  below the (flat) threshold level that attracts the relaxation of initial states at $T'\gg T_d$. Several later papers improved the analysis and confirmed the result just described~\cite{Capone2006,Sun2012,Dembo2019}. In particular, in Ref.~\cite{Folena2019}, equilibrium initial conditions at $T'\in [T_d,T_{\rm onset}]$ were considered and, very surprisingly, memory of the initial conditions after quenches to very low temperatures was observed in the numerical solutions of the Schwinger-Dyson equations. More details on this and other peculiar features of the metastable and dynamic properties of the mixed model are given in the Background Subsection~\ref{subsec:background}.

In this paper we first show that chaos in temperature, in a sense that we will make precise later, is present in the low temperature regime $T<T_d$ (and not only below $T_s$) in the mixed model. We then introduce and study a constrained free energy density function, of Thouless-Anderson-Palmer (TAP) \cite{Thouless1977} type but with new conditions, that allows one to identify a possible origin of the differences in the quenched dynamic behaviour of the mixed and pure spherical models. We also set the stage for a generalisation of the dynamic approach to follow the evolution of equilibrium initial conditions in more detail than done so far. 

The paper is structured as follows. In Section~\ref{sec:model} we present the model and we recall how its stochastic dynamics are described via a Langevin process. Section.~\ref{sec: temperature chaos} focuses on temperature chaos captured by the  Franz-Parisi (FP) potential and the TAP free energy. In Sec.~\ref{sec: A TAP-like approach to quenches} we introduce the  constrained TAP free energy approach. Sections~\ref{sec:Application to the mixed $p$-spin model} and~\ref{se: an exact approach for the constrained free energy} are devoted to the derivation of our results and their discussion, also in connection with predictions from the use of the FP potential. A concluding Section closes the paper. In four appendices we present some properties of the unconstrained TAP free energy landscape and we provide details on the derivation of the constrained one.

\section{The model} 
\label{sec:model}

In this Section we introduce the model
and we recall some of its most relevant properties.

\subsection{Definitions} 
\label{subsec:definitions}

We study a disordered spherical spin model with Hamiltonian equal to the 
sum of two $p$-spin terms:
\begin{eqnarray}
    H_J[\{s_i\}]
    &=&H_{p_1}[\{s_i\}]+H_{p_2}[\{s_i\}]=-\sum_{i_1<\dots<i_{p_1}}J_{i_1\dots i_{p_1}} s_{i_1}\dots s_{i_{p_1}}-\sum_{i_1<\dots<i_{p_2}}J_{i_1 \dots i_{p_2}}s_{i_1}\dots s_{i_{p_2}}
    \; . 
    \label{eq:Hamiltonian}
\end{eqnarray}
The $i=1,\dots,N$ spin variables are real and continuous, $-\infty < s_i < \infty$,
and they are globally constrained to satisfy
\begin{equation}
\frac{1}{N} \sum_{i=1}^N s^2_i = 1
\; . 
\label{eq:spherical-constraint}
\end{equation}
Quenched disorder is introduced by the interaction constants 
$J_{i_1 \dots i_{p_\ell}}$ that are {\it independent} random variables 
taken from two Gaussian distributions with mean and variance
\begin{equation}
{\rm I\!E} [J_{i_1 \dots i_{p_\ell}}]=0 \qquad\mathrm{and}\qquad {\rm I\!E} [{J^2_{i_1 \dots i_{p_\ell}}}]=\frac{{J_{p_\ell}^2}p_\ell!}{2{N}^{p_\ell-1}}
\end{equation} 
where $J_{p_\ell}>0$ and $\ell=1,2$.
The random exchanges induce correlations between the Hamiltonian (\ref{eq:Hamiltonian}) evaluated on two different spin configurations
$\{s_i\}$ and $\{s'_i\}$. Defining their overlap 
\begin{equation}
C_{ss'} \equiv \frac{1}{N} \sum_{i=1}^N s_i s'_i
\end{equation}
one has
\begin{eqnarray}
\nu(C_{ss'})
\equiv
{\rm I\!E}[H_J[\{s_i\}]H_J[\{s'_i\}]] 
=
\frac{J_{p_1}^2}{2} C_{ss'}^{p_1}
+
\frac{J_{p_2}^2}{2} C_{ss'}^{p_2}
\; . 
\label{eq:random-pot-corr}
\end{eqnarray}
For later convenience we called the expectation value $\nu$.

The stochastic dynamics are governed by overdamped Langevin equations
\begin{equation}
    \eta \frac{d s_i(t)}{dt} = 
    - \frac{\delta H_J[\{s_j(t)\}]}{\delta s_i(t)} - \mu(t) s_i(t) + \xi_i(t)
    \; , 
    \label{eq:Langevin}
\end{equation}
with $\mu(t)$ a Lagrange multiplier that imposes the spherical constraint (\ref{eq:spherical-constraint}) all along the evolution, and $\xi_i(t)$
a time dependent Gaussian random force with zero mean and delta-correlations:
\begin{equation}
    \langle \xi_i(t) \rangle = 0 
    \; ,
    \quad
    \langle \xi_i(t) \xi_j(t')\rangle = 2 \eta k_BT \delta(t-t') \delta_{ij}
    \; . 
\end{equation}
The evolution starts from initial conditions
$\{s_i(0)\}$ that are chosen with different criteria. The ones most commonly used are in equilibrium at temperature $T'$. In the infinite temperature limit, $T'\to\infty$, their statistics is mimicked with a flat probability distribution. At finite temperature, $T'<+\infty$, the disordered dependent Gibbs-Boltzmann weight at $T'$ is used to sample $\{s_i(0)\}$. 

In the thermodynamic limit the correlation function $C(t,t')$ and the linear response function $R(t,t')$ are the main observables that describe the dynamics of the system. The first one consists in the average (over the thermal noise and initial conditions denoted with angular brackets, the disorder indicated with ${\mathbb E}$, and the whole system) overlap of a spin $s_i$ taken at two times $t$ and $t'$ strictly larger than the initial one, that hereafter we set to zero:
\begin{equation}
C(t,t')=\frac{1}{N}\sum_i {\rm I\!E}\Big[\langle s_i (t) s_i (t') \rangle\Big]
\; .
\label{eq:Cttp}
\end{equation}
We choose to distinguish this `late times' function from the correlation between the initial configuration $\{s_i(t=0)\}$ and the configuration at a later time  $\{s_i(t>0)\}$:
\begin{equation}
C(t,0)=\frac{1}{N}\sum_i {\rm I\!E}\Big[\langle s_i (t) s_i (0) \rangle\Big]
\; .
\label{eq:Ct0}
\end{equation}

The response function is calculated from the variation between the evolution, on average, of a given spin $s_i$ with the addition of a magnetic field $h_i(t)$, such that the forces are shifted by $-\delta_{s_i(t)} H[\{ s_j(t)\}] \rightarrow -\delta_{s_i(t)} H[\{ s_j(t)\}]+h_i(t)$, and the free one. More formally, the linear response function can be written as
\begin{equation}
R(t,t')=\frac{1}{N}\sum_i {\rm I\!E}\Big[  \delta_{h_i(t')} \langle s_i (t)\rangle_h \Big]\Bigr|_{h_i(t)=0} \;.
\end{equation}

In the following we set the units such that $\eta=k_B=1$.
It is known that this model has different static, metastable and dynamic behaviour depending on whether one of the two $p$ parameters takes the value $2$ or not, see Ref.~\cite{Crisanti2006} for details. In the following study we will choose the convention $p_1<p_2$
and we will focus on $2<p_1$.

\subsection{Background} 
\label{subsec:background}

As we have already written in the Introduction, these models have an equilibrium phase transition at a temperature $T_s$ determined from, for example, the analysis of the symmetry breaking properties in the replica calculation of the thermodynamic free energy. The replica structure goes from being symmetric above $T_s$ to one step symmetry breaking (RSB) below it, indicating the presence of a glassy equilibrium phase at low temperatures. The transition is discontinuous in the sense that the order parameter jumps but second order thermodynamically. There is no Gardner temperature below which a full RSB solution would be needed in this model.  The equilibrium phase diagram is discussed in detail in App. C in Ref.~\cite{Crisanti2006}.

The relaxation dynamics from random, infinite temperature, initial conditions, face the impossibility to equilibrate below a temperature $T_d$ ($>T_s$). Still, the correlation function with the initial condition and the two-time one for widely separated times approach zero, when times are taken to diverge after the thermodynamic limit, in the whole post-quench temperature range of variation. Below $T_d$, this complete decorrelation gives rise to the so-called {\it weak ergodicity breaking} scenario~\cite{Bouchaud1992,Cugliandolo1995}. The relaxation approaches a flat region of phase space named the threshold~\cite{Cugliandolo1993}. The dynamic transition at $T_d$ is also discontinuous. These conclusions can be extracted from Ref.~\cite{Cugliandolo1996} since the mixed model is a special case of the ones studied in this reference. The dynamic transition line can also be found with a replica study in which marginality is imposed, see the App. C in Ref.~\cite{Crisanti2006} for the development of this approach and Fig. 17 in this reference for the phase diagram of the mixed model with $p_1=3$ and $p_2=4$.

The static phases and phase transitions, and the dynamic properties after quenches from infinite temperature, just described are in complete analogy with the ones of the pure $p$-spin spherical model.

Thouless, Anderson \& Palmer (TAP)~\cite{Thouless1977} introduced a formalism that allows one to define and investigate a free energy landscape that is a function of all relevant order parameters and thus access metastable states of all kinds. This approach extends Landau's to disordered systems.
For the disordered spin models we are dealing with, the order parameters are the local magnetisations, $\langle s_i\rangle = m_i$, and they are order $N$ in number. In pure $p$-spin models, the TAP free energy landscape is complex but relatively simple at the same time~\cite{Kurchan1993,Crisanti1995}. It starts having a complex structure, with stationary points that are associated to metastable states, at temperatures that are well above $T_s$. But these states are organised in such a way that they neither cross, merge nor bifurcate; therefore, once one of them is identified at, for example, zero temperature, it can be followed in temperature until it disappears at its spinodal.
Pure $p$-spin models have, in a finite window of temperatures above $T_s$, an exponentially large number of non-trivial states with, e.g., different and non zero local magnetisations, $\{m_i^{\alpha}\}$ with $\alpha$ the state identification,
that combine to yield paramagnetic global properties. De Dominicis and Young~\cite{deDominicis1983}
showed that proper equilibrium averages can be recovered from the average of the value of the selected observable, $O$, in each of these states, $O(\{m_i^\alpha\})$, weighted with a Boltzmann probability factor $e^{-\beta N f(\{m_i^\alpha\})}/Z$, and summed over all $\{m_i^\alpha\}$.  In the development of this calculation the number of metastable states with the same TAP free energy
density, ${\cal N}(f)$, plays a crucial role. Indeed, the sum over $\{m_i^\alpha\}$ is transformed into an integral over $f$, and the complexity or configurational entropy, that is to say, the logarithm of their number, $\Sigma(f) = \ln {\cal N}(f,T)$, intervenes in the statistical weight that is modified, in the continuum limit, to be $\exp[-\beta N (f-T \Sigma(f,T))]$.

This nice structure is partly due to the homogeneity of the monomial potential of the pure models and it is partially lost in the mixed problems, that present {\it temperature chaos}~\cite{Rizzo2006}. 

Barrat {\it et al.}~\cite{Barrat1997} calculated the Franz-Parisi (FP) effective potential~\cite{Franz1995} as an alternative way to observe the bifurcation of metastable  states in the mixed model.
The FP potential is the Legendre transform of the free energy of the system under a local field proportional to 
a particular equilibrium configuration at a chosen temperature $T'$. The dependence on the strength of the 
local field, say $\epsilon$, is exchanged, under the Legendre transform,
into a dependence on the overlap between the reference configuration 
and the ones at the working temperature. The FP potential is, therefore, the free energy cost to keep a 
system in equilibrium at temperature $T$ at a fixed overlap with a generic equilibrium 
configuration at another temperature $T'$. The need to break replica symmetry in the mixed model to calculate this potential 
below another characteristic temperature  $0<T_{\rm RSB}(T')<T'$
was interpreted as a signature of the multifurcation of metastable states below this same temperature.

In the same paper, Barrat {\it et al.}~\cite{Barrat1997} derived and performed a first study of the Schwinger-Dyson dynamic equations for the disorder averaged
model quenched from equilibrium at a temperature $T_s< T'<T_d$ to a lower temperature $T<T'$. The average over the equilibrium initial conditions was dealt with using the replica trick, as pioneered in Ref.~\cite{Houghton83} and, since
$T'>T_s$, no replica symmetry breaking was used. Nevertheless, in this range of temperatures, a complex TAP free energy landscape already exists (as discussed in the third paragraph in this Section). Therefore, the initial configurations drawn with the Gibbs-Boltzmann measure are interpreted as being within one non-trivial TAP state with non-zero values of the local magnetisations $\{m_i^\alpha \neq 0\}$ that, however, are averaged over in this calculation and are not individually accessed. The authors showed that above the 
temperature  $T_{\rm RSB}(T')$ the dynamics occur as in equilibrium and the correlation with the initial condition does not approach zero but a value consistent with the state following interpretation. However, below $T_{\rm RSB}(T')$ these solutions no longer exist and
the authors conjectured that the dynamics age 
forever with the very unusual feature of keeping a memory of the initial condition, via a non-zero asymptotic 
value of the correlation function 
$C(t,0)$ 
({\it strong ergodicity breaking}). The picture developed in this paper was later 
confirmed in~\cite{Sun2012} where a planting procedure was used to 
generate the initial conditions and the adiabatic state following method~\cite{Krzakala2010} was applied.


Next, Capone {\it et al.}~\cite{Capone2006} studied the Schwinger-Dyson equations for equilibrium initial conditions (also imposed with a replica calculation) in more detail
than done in Ref.~\cite{Barrat}. On the one 
hand, they confirmed the results of Barrat {\it et al.}~\cite{Barrat1997} with usual restrained equilibrium state following above $T_{\rm RSB}(T')$ and ageing below this temperature
taking place in a marginal manifold (supposedly the one in which the initial state opens up) that lies {\it below} the threshold one approached with quenches from $T' \rightarrow +\infty$. However, they also realised that the asymptotic equations derived with an ageing {\it Ansatz} that fix, for example, the varios long-time limit values of the correlation, {\it do not have solution} below another characteristic temperature $T_c(T')<T_{\rm RSB}(T')$. (These equations are the same that fix the parameters $q_o$, $q$ and $x$ in the 1RSB calculation of the 
FP potential that, therefore, do not have solution either below $T_c(T')$.) The authors complemented the dynamic analysis with 
a static one in which they calculated a {\it constrained complexity}, defined as the number of states at temperature $T$ with given free energy density and
overlap with {\it all} reference equilibrium states at $T'$.  The temperature 
$T_{\rm RSB}(T')$ was then associated with the one at which this constrained complexity vanishes. 

Several new features of the quench dynamics of the mixed model have recently been shown with a numerical integration of the Schwinger-Dyson equations ~\cite{Folena2019}. The authors identified a temperature $T_{\rm onset}$, higher than the usual dynamical temperature $T_d$, below which the system memorises the initial condition
when instantaneously quenched to a sufficiently low temperature. 
They have also shown that the system can go through an ageing regime where the description used for the pure $p$-spin case fails. 
In fact, the marginal states reached through this ageing dynamics have a non-zero overlap with the initial condition, and the usual analytical {\it Ansatz} with weak long term memory and weak ergodicity breaking features used to describe ageing regimes~\cite{Cugliandolo1993} does not fit the simulations because of this fact.

 With the aim of clarifying the origin of the unexpected behavior found in the references cited above~\cite{Barrat1997,Capone2006,Sun2012,Folena2019}, we here revisit the TAP approach by using new constraints, \`a la FP. The idea is to keep track of the individual TAP states that contribute to the equilibrium measure at $T'$. These, identified with the states where the initial conditions are located, we claim, should have a distinctive dynamic evolution.
 
 In order to clarify followig discussions and to set orders of magnitude the values of $T_s$ and $T_d$ for $p_1=3$, $p_2=4$ and $J_{p_1}=J_{p_2}=1$ are
 \begin{equation}
 \label{eq: temp. charac.}
     T_s\approx 0.762 \quad \text{and}\quad T_d\approx 0.805\; . 
 \end{equation}
 We recall that we consider $T' \in [T_s,T_d]$ and $T_{\rm RSB}(T')$ is a function of $T'$ that varies from $0$ to $T_d$.
 
 \section{Temperature chaos}
\label{sec: temperature chaos}

Let us consider, as in Refs.~\cite{Barrat1997,Capone2006,Folena2019}, an 
equilibrated system at $T'$, described by the Gibbs-Boltzmann distribution $P[\{s_i\}]\propto \exp\big(-\beta' H_J[\{s_i\}]\big)$.

The unrestrained TAP analysis shows that the equilibrium measure in the temperature window $T'\in [T_s;T_d]$
is dominated by an ensemble of non-trivial TAP states with $\{m_i \neq 0\}$, in the sense
that they are the ones that dominate the measure $Z^{-1}(\beta') \, 
\exp[-\beta' N(f-T' \Sigma(f,T'))]$
with $\Sigma(f,T')$ the complexity calculated in Refs.~\cite{Rieger1992,Rizzo2006}.
TAP states are fully parametrised by their overlap $q = N^{-1} \sum_i m_i^2$ and the adimensional energy densities $\varepsilon_{p_1}= N^{-1} J_{p_1}^{-1} q^{-p_1/2} H_{p_1}[\{m_i\}]$ 
and $\varepsilon_{p_2}=N^{-1}J_{p_2}^{-1} q^{-p_2/2} H_{p_2}[\{m_i\}]$, see Eq.~(\ref{eq:fTAP}) and App.~\ref{sec:app_TAP}. 
Therefore, at each temperature $T'$, 
the measure above is dominated by TAP states with optimised values of $q_{eq}$, 
$\varepsilon_{p_1, eq}$ and $\varepsilon_{p_2, eq}$ 
given by
\begin{eqnarray}
\label{eq: equilibrium}
    \frac{q_{eq}}{1-q_{eq}} &=& 
    \frac{p_1\beta^2 J_{p_1}^2}{2} q_{eq}^{p_1-1}+\frac{p_2 \beta^2 J_{p_2}^2}{2} q_{eq}^{p_2-1}
    \; ,
\\
    \varepsilon_{p_{\ell,eq}}&=&-\frac{\beta J_{p_\ell}}{2}\Big[p_\ell q_{eq}^{\frac{p_\ell}{2}-1}-(p_\ell-1)
    q_{eq}^{\frac{p_\ell}{2}} \Big]
\qquad \mbox{for} \quad \ell=1,2
\; . 
\end{eqnarray}
This is similar to what happens in the pure model though with an extra `parameter' $\varepsilon_{p_2,eq}$. In the following we will consider that a TAP state is ``followed" after a change in temperature whenever the temperature change induces a simple homothetic transformation (global rescaling or homogeneous dilation) of the magnetisations configuration. In other words if the magnetisations of the state at the new temperature are simply  $m_i \rightarrow \alpha \, m_i $. Whenever this property is not verified it is straightforward to see that $\varepsilon_{p_1,eq}$ and $\varepsilon_{p_2,eq}$ change with temperature, in this case we will talk about chaotic behavior.

The FP potential is well adapted to study the equilibrium
behaviour of a system at temperature $T$, constrained to have a given overlap with itself when in equilibrium at another temperature $T'$. Moreover, 
the results for the relevant parameters found with the FP match the asymptotic overlaps $q_o$, $q$ and the energy density derived with a dynamic approach in which the system is initialised in equilibrium at $T'$ and evolved at a different temperature $T$.
In fact, both FP and dynamics approaches
yield parameters $q_o$ and $q$ determined by the set of equations
\begin{eqnarray}
\label{eq: dynamics (1)}
    q_o^2 &=&
    q-(1-q)^2 \Big[\frac{p_1 J_{p_1}^2\beta^2}{2}q^{p_1-1}+\frac{p_2 J_{p_2}^2\beta^2}{2}q^{p_2-1}\Big]
    \; , 
\\
\label{eq: dynamics (2)}
        \frac{1}{1-q} &=&
        \frac{p_1 J_{p_1}^2\beta'\beta}{2} q_o^{{p_1}-2}+\frac{p_2 J_{p_2}^2\beta'\beta}{2} q_o^{{p_2}-2}
        \; , 
\end{eqnarray}
as long as $T \in [T_{\rm RSB}(T'),T']$.
In the FP calculation~\cite{Barrat1997}, 
$q$ is the order parameter of the constrained system and $q_o$ its overlap with the equilibrated system at $T'$. In the dynamic calculation~\cite{Barrat1997,Capone2006}, $q_o = \lim_{t\to\infty} C(t,0)$, see Eq.~(\ref{eq:Ct0}), while $q=\lim_{t\to\infty} 
\lim_{t'\to\infty} C(t,t')$, see Eq.~(\ref{eq:Cttp}).
Both approaches exhibit the same transition temperature $T_{\rm RSB}(T')$ which has been interpreted as the start of an ageing regime where the system
approaches marginal states
with order parameter determined by a different equation
\begin{equation}
\label{eq:q_marginal}
    T^2 = \nu''(q_{\rm marg}) (1-q_{\rm marg})^2  
 \; .
\end{equation}

The comparison of results obtained with the FP 
potential and the usual unconstrained TAP free energy can already show that there should be temperature chaos in the temperature interval $[T_{\rm RSB}(T'),T']$. We justify this claim as follows.

The energy density of the system at temperature $T$
can be obtained with the two approaches and 
compared. On the one hand, the system's 
energy density, in the $[T_{\rm RSB}(T'),T']$ interval, calculated with the FP potential is
\begin{eqnarray}
\label{eq: quench dynamics(1)}
       e_{\rm FP}
       &=&
       -\frac{\beta'}{2}(J_{p_1}^2 q_o^{p_1}+J_{p_2}^2 q_o^{p_2})-\frac{\beta}{2}\Big[J_{p_1}^2 (1-q^{p_1})+J_{p_2}^2 (1-q^{p_2})\Big] 
\end{eqnarray}
(this expression can be read from Eq.~(27) in Ref.~\cite{Barrat} setting the last term to zero and making the necessary changes of names of variables, $\tilde p = q_0$ and $q_1=q$.) Using Eqs.~(\ref{eq: dynamics (1)}) and (\ref{eq: dynamics (2)}) to replace $q_o$ and $q$, one can readily get the dependence of $e_{\rm FP}$ on $T$ and $T'$.

\begin{figure}[h]
    \centering
    \includegraphics[width=0.55\textwidth]{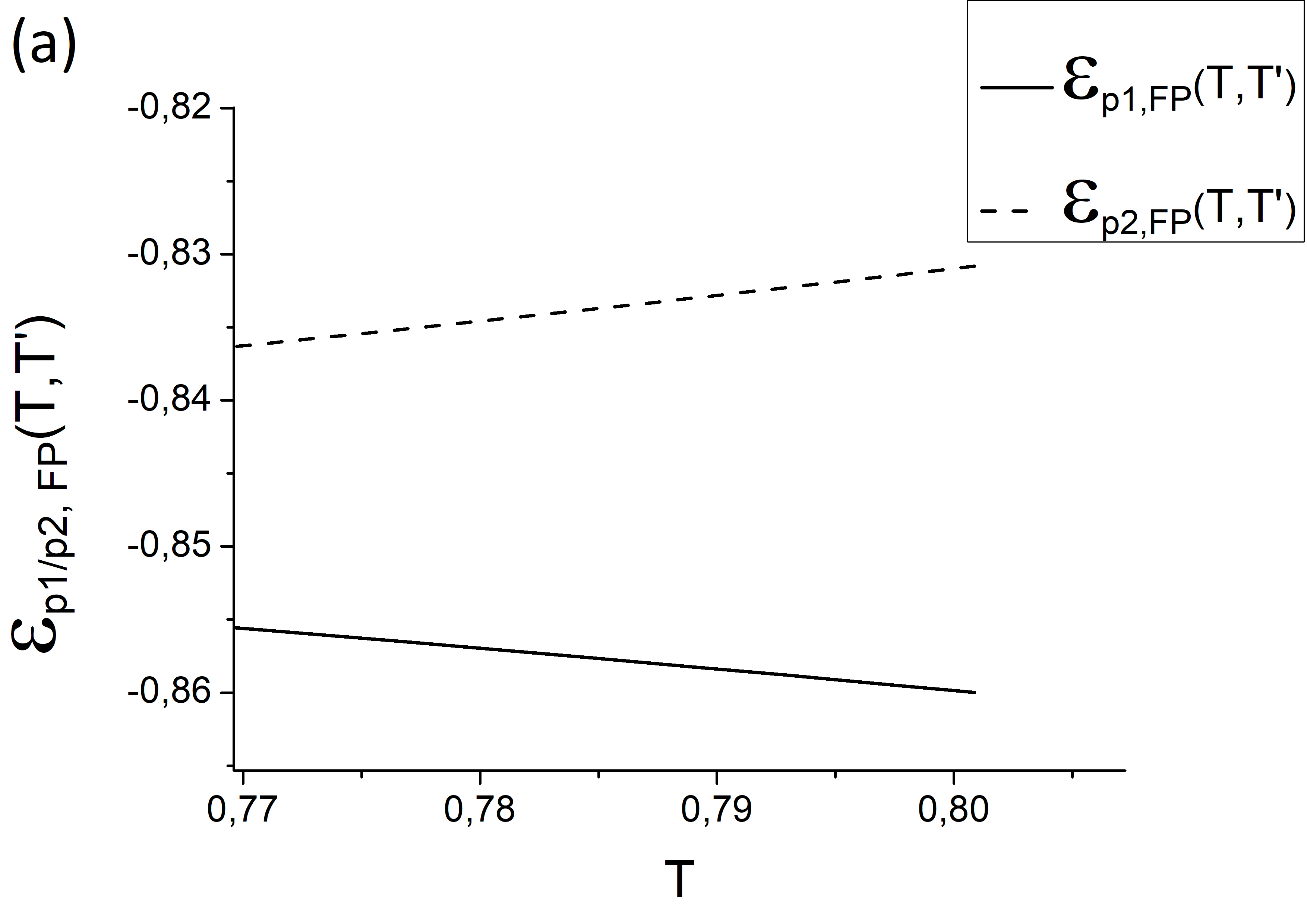}
    \quad\quad
    \includegraphics[width=0.55\textwidth]{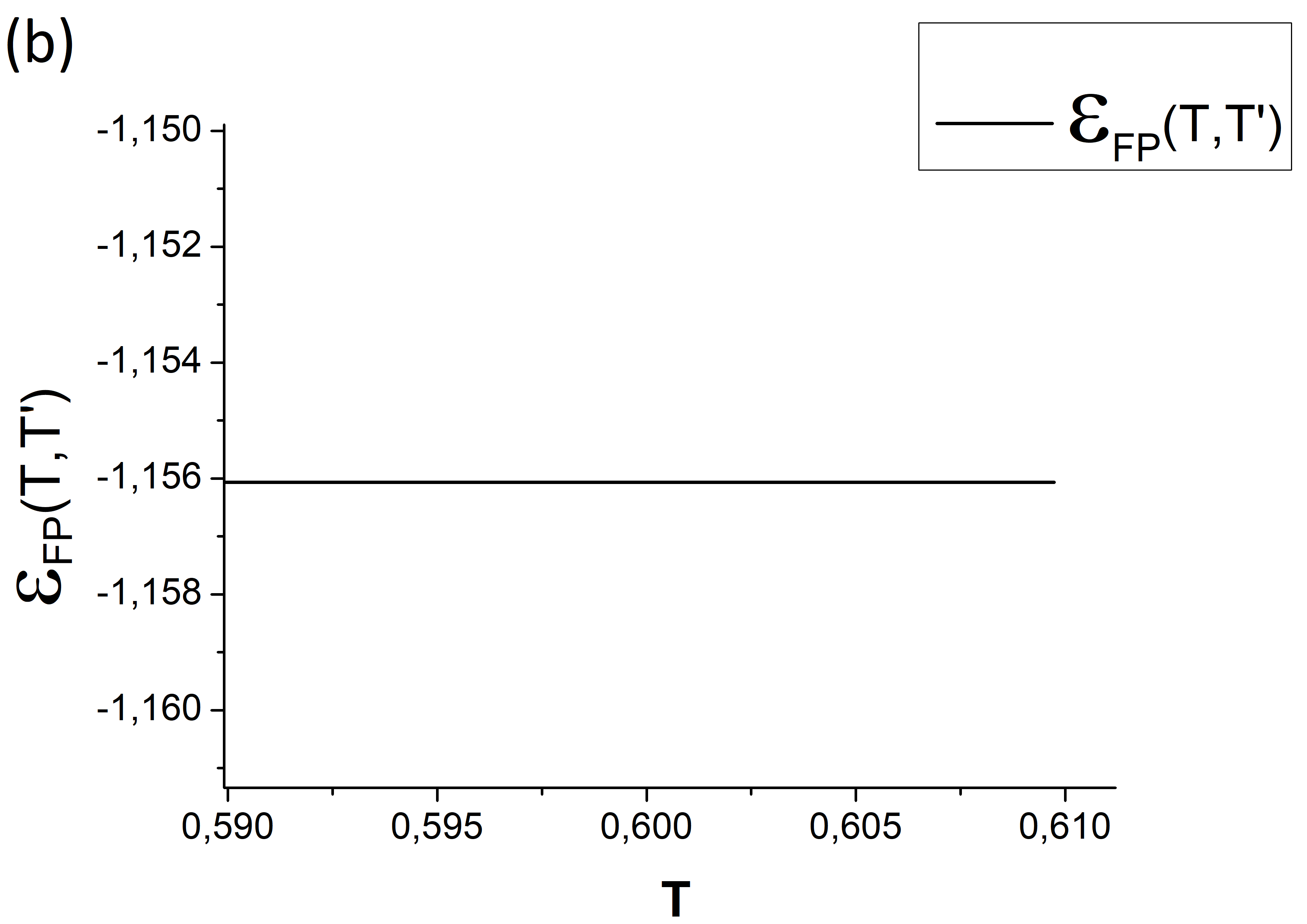}
    \caption{ The adimensionnal energy densities are derived with the FP potential as explained at the end of Sec.~\ref{sec: temperature chaos} (a proxy for a system equilibrated at $T'=1/\beta'$ and quenched to a bath temperature $T=1/\beta$). Panel (a) focuses on a mixed $p$-spin model with $T'\approx 0.801$, $p_1=3$, $p_2=4$ and $J_{p_1}=J_{p_2}=1$ ($T_s$ and $T_d$ are recalled in Eq.~(\ref{eq: temp. charac.})). Panel (b) focuses on the pure $p$-spin model with $p=3$, $T'\approx 0.609$ ($T_s\approx0.586$ and $T_d\approx0.612$), $J=1$ and we retrieve the state following behavior as $\varepsilon_{FP}(T,T')$ is constant for all temperatures $T$.
   }
    \label{fig:energy_variation}
\end{figure}

\begin{figure}[h]
    \centering
    \includegraphics[width=0.6\textwidth]{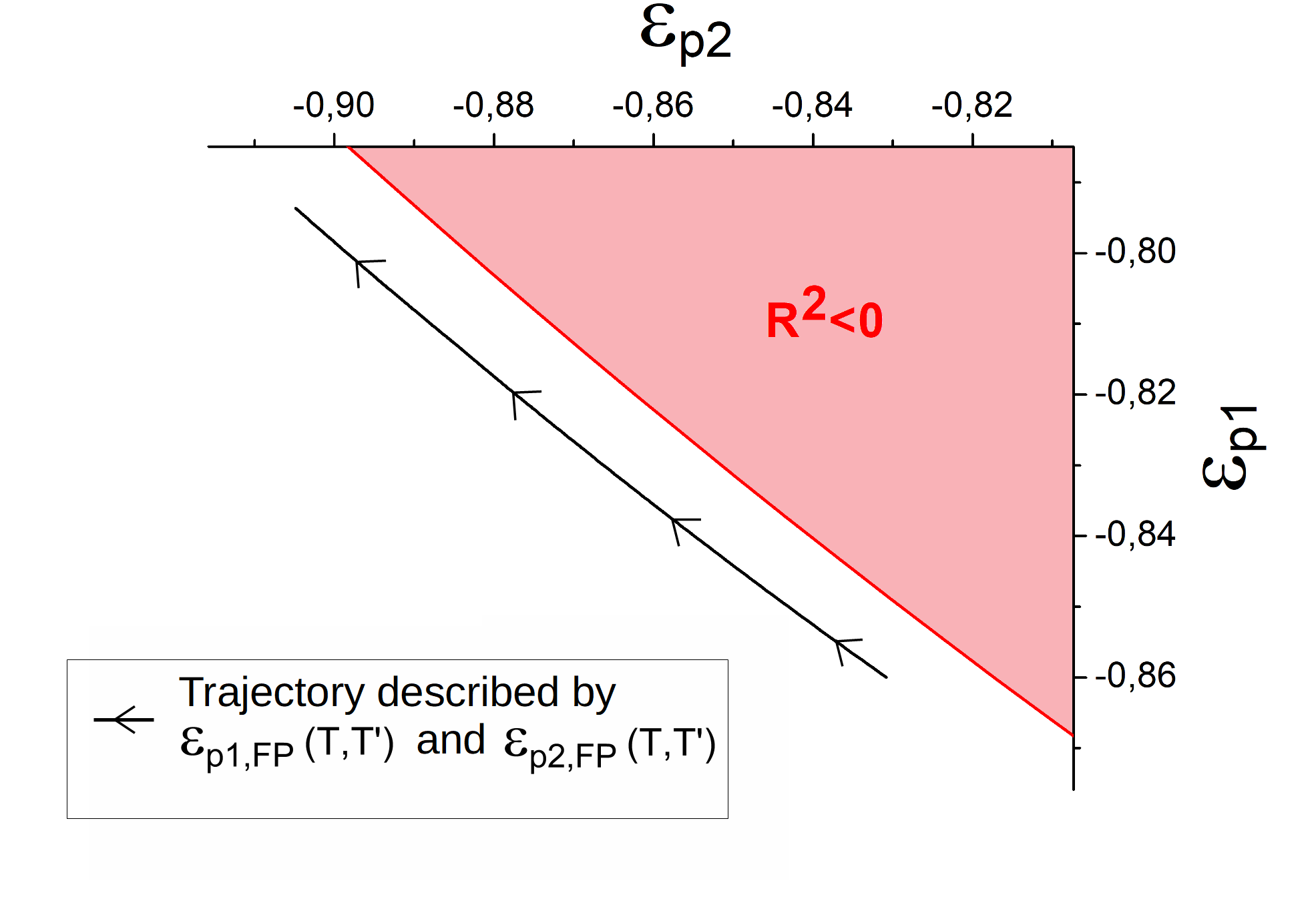}
    \caption{We reproduced the energy density diagram for metastable TAP states -described in App.\ref{sec:app_TAP}- with the exclusion zone bounded by the limit states. On top of it we added the trajectory described by the adimensionnal energy densities $\varepsilon_{p_1,\rm FP}(T,T')$ and $\varepsilon_{p_2,\rm FP}(T,T')$. It emphasises the chaos in temperature when a system is equilibrated at a bath temperature $T$ and constrained with a system at $T'$. The trajectory was obtained for $p_1=3$, $p_2=4$, $J_{p_1}=J_{p_2}=1$ and $T'\approx 0.801$. It starts at $T=T'$ and following the direction of the arrows the temperature $T$ gets lower and lower.}
    \label{fig:energy_variation(2)}
\end{figure}

On the other hand, the energy density derived from the TAP free energy is
       \begin{eqnarray}
        e_{\rm TAP}
        &=&  q^{p_1/2} J_{p_1} 
        \varepsilon_{p_1}+  q^{p_2/2} J_{p_2}\varepsilon_{p_2}-\frac{ \beta J_{p_1}^2}{2}\Big[(p_1-1)q^{p_1}-{p_1}q^{p_1-1}+1\Big] -\frac{ \beta J_{p_2}^2}{2}\Big[(p_2-1)q^{p_2}-{p_2}q^{p_2-1}+1\Big]  \; 
        \label{eq:eTAP}
\end{eqnarray}
with $q$ the order parameter $q=\frac{1}{N}\sum_i m_i^2$ and $\{m_i\}$ the local magnetisations in a TAP state. We eliminated the labels $eq$ used in Eqs.~(\ref{eq: dynamics (1)}) and (\ref{eq: dynamics (2)}) for simplicity. The equations that fix the $N$
local magnetisations, $\partial f_{\rm TAP}/\partial m_i =0$, can be multiplied by $m_i$ and summed over $i$
to yield an extra equation that relates $q$ to $\varepsilon_{p_1}$ and $\varepsilon_{p_2}$:
\begin{eqnarray}
\label{eq: quench dynamics(2)}
       0=1+ \beta \frac{1-q}{q} \sum_{\ell=1}^2  J_{p_\ell}
       p_\ell q^{p_\ell/2} \varepsilon_{p_\ell} +
       \beta^2 (1-q)^2 
       \sum_{\ell=1}^2
      J_{p_\ell}^2 \, \frac{p_\ell(p_\ell-1)}{2} \,  q^{p_\ell-2} \; .
\end{eqnarray}
We use this condition to obtain $\varepsilon_{p_1}$ as a function of 
$q, \varepsilon_{p_2}$ and $T$ that we replace in Eq.~(\ref{eq:eTAP}) 
and thus rewrite the TAP energy density in the form 
$e_{\rm TAP}(q,\varepsilon_{p_2}, T)$. 
If we now require that $q$ (and $q_o$) 
be determined by Eqs.~(\ref{eq: dynamics (1)}) and (\ref{eq: dynamics (2)}) we can rewrite 
the TAP energy density in a new form that is $e_{\rm TAP}(\epsilon_{p_2}, T, T')$.
If the systems at temperature $T$ described by the TAP and FP approaches were the same, the FP and TAP energies should coincide and the condition
$e_{\rm FP}(T,T')=e_{\rm TAP}(\varepsilon_{p_2},T,T')$ verified. This gives an equation that determines $\varepsilon_{p_2}(T,T')$. In the following we will call 
$\varepsilon_{p_2,{\rm FP}}(T,T')$ the adimensional energy density obtained through this method. One can check numerically, see Figs.~\ref{fig:energy_variation} and \ref{fig:energy_variation(2)},  that in the mixed model $\varepsilon_{p_2, {\rm FP}}$ thus obtained depends on both temperatures $T$ and $T'$. 
Hence, for any temperature $T \in [T_{\rm RSB}(T'),T'[$ the constrained system shifts away from the original TAP state. On the contrary, in the pure model, the same construction yields a $T$-independent energy density $\varepsilon_p$, indicating that there is no chaos in temperature in this case.

\section{A constrained TAP free energy density}
\label{sec: A TAP-like approach to quenches}

The TAP approach consists in probing the local minima of the (rough) free energy landscape with respect to the local magnetisations $\langle s_i \rangle=m_i$, where the angular brackets denote a static statistical average. This description allows one to reach an understanding of metastability in disordered mean-field models. Moreover, it enabled one to recover equilibrium results, originally derived with the replica trick~\cite{Crisanti2006}, and to grasp the outcome of the relaxation dynamics following quench protocols from disordered~\cite{Biroli1999,Cugliandolo1993}
and metastable initial conditions~\cite{Barrat1996,Biroli1999,Franz1995} in the pure $p$-spin model. 

Different methods to obtain the TAP free energy and the ensuing TAP equations have been developed throughout the years~\cite{Biroli1999,Crisanti1995,Mezard1987,Rieger1992,Thouless1977}. One can cite, for example, the cavity method, the diagrammatic expansion of the free energy or the historical derivation by Thouless, Anderson and Palmer~\cite{Thouless1977}. We use here the proof based on the Legendre transform of the thermodynamic free energy, first introduced by Georges and Yedidia \cite{Georges1991}.

\subsection{Justification of the approach and definition of the free energy}
\label{sec: justification of the approach}

Let us take a vector in the $N$ dimensional phase space  with components $\{v_i\}$. It could be given by the ensemble of 
local magnetisations $\{ v_i = m_i^{\sigma} \}$ that characterise a TAP state at temperature $T'$, where $\sigma$ is the label that identifies the TAP state chosen, or it could be just a generic $N$-dimensional vector. 

We require that the (thermal averaged) overlap between a configuration
$\{s_i\}$ and this vector be 
\begin{equation}
  \sum_i \langle s_i \rangle v_i =Nq_o
\; . 
\end{equation}
In this section we will compute the free energy of a system, at temperature $T=1/\beta$, when the configurations are constrained to have, on average, 
overlap $q_o$ with the reference configuration defined by 
$\{ v_i\}$.
More explicitly, we calculate the constrained free energy
 \begin{eqnarray}
 \label{eq: free energy def}
    -\beta F_J [\beta,q_o,\{v_i\},l]&=&\ln \Bigg\{  {\rm Tr}_{\{s_i\}}\Big[e^{-\beta H_J[\{s_i\}]-\frac{\lambda}{2}\sum_i (s_i^2-l)-h\sum_i(s_i v_i-q_o)}\Big]  \Bigg\}=\ln Z_J[\beta,q_o,\{v_i\},l]\\
    &=&-\beta F_J^\star[\beta,\{h v_i\},\lambda]+\frac{N \lambda l}{2}+N h q_o \nonumber
 \end{eqnarray}
 up to order $O(N)$, with
 \begin{eqnarray}
   \partial_\lambda \big(-\beta F_J[\beta,q_o,\{v_i\},l]\big)    &=& 0 \implies \sum_i \langle s_i^2 \rangle= N l \; ,
\\
       \partial_h \big(-\beta F_J[\beta,q_o,\{v_i\},l]\big)    
       &=&    
       0 \implies \sum_i \langle s_i \rangle v_i=Nq_o \; ,
\end{eqnarray}
or in another fashion
\begin{equation}
\label{eq: constraining the free energy}
       \partial_\lambda \big(-\beta F_J^\star[\beta,\{h v_i\},\lambda]\big)    
       =  
       \frac{N l}{2} \implies \sum_i \langle s_i^2 \rangle= N l \; ,
\end{equation}
\begin{equation}
\label{eq: constraining the free energy(2)}
       \partial_h \big(-\beta F_J^\star[\beta,\{h v_i\},\lambda]\big)    
       =   
       Nq_o \;\implies \sum_i \langle s_i \rangle v_i=Nq_o \; .
 \end{equation}
The function $F_J^\star[\beta,\{h v_i\},\lambda]$ is the Legendre transform of $F_J[\beta,q_o,\{v_i\},l]$. Moreover the parameter $\lambda$ enforces the spherical constraint while $h$ fixes the global overlap with the reference state $\{v_i\}$. 
 
In the end the free energy defined in this way has to be extremised with respect to $q_o$. Two arguments can be offered to justify this statement. The first one consists in requiring that the equilibrium properties of the system be described by the thermodynamic free energy
\begin{equation}
     -\beta F_J[\beta,l]=\ln \Bigg\{  {\rm Tr}_{\{s_i\}}\Big[e^{-\beta H_J[\{s_i\}]-\frac{\lambda}{2}\sum_i (s_i^2-l)}\Big]  \Bigg\}=\ln Z_J[\beta,l]
     \label{eq:free energy}
 \end{equation}
 where the parameter $\lambda$ still enforces the spherical constraint. Thus, if the constrained free energy 
 $F_J[\beta,q_o,\{v_i\},l]$ 
 is made extreme for $q_o=\hat{q}_o$, one has
 \begin{eqnarray}
      \partial_{q_o} \big(-\beta F_J[\beta,q_o,\{v_i\},l]\big)\Bigr|_{q_o=\hat{q}_o} =0 \implies h\Bigr|_{q_o=\hat{q}_o}=0
 \end{eqnarray}
and one recovers 
 \begin{eqnarray}
F_J[\beta,\hat{q}_o,\{v_i\},l]=\ln \Bigg\{  {\rm Tr}_{\{s_i\}}\Big[e^{-\beta H[\{s_i\}]-\frac{\lambda}{2}\sum_i (s_i^2-l)}\Big]  \Bigg\}=-\beta F_J[\beta,l]
 \end{eqnarray}
 with
 \begin{equation}
 \sum_i \langle s_i \rangle v_i =N \hat{q}_o
 \; . 
 \end{equation}
The second argument is based on the usual saddle point approximations performed for extensive quantities.
Indeed, the thermodynamic free energy  can be rewritten as 
\begin{eqnarray}
     -\beta F_J[\beta,l]&=&
      \ln \Bigg\{ {\rm Tr}_{\{s_i\}}\Big[ \int dq_o \; e^{-\beta H_J[\{s_i\}]-\frac{\lambda}{2}\sum_i (s_i^2-l)}\Big] \delta\big(\sum_i s_i v_i-q_o\big) \Bigg\}\\
      &=&
      \ln \Bigg\{ {\rm Tr}_{\{s_i\}}\Big[ \int dh \; dq_o \; e^{-\beta H_J[\{s_i\}]-\frac{\lambda}{2}\sum_i (s_i^2-l)+h(\sum_i  s_i v_i -q_o)}\Big]  \Bigg\} 
     \nonumber\\
     &=& 
     \ln \int dh dq_o \; e^{-\beta N f_J[\beta, \, h \!, \, q_o, \{v_i\},l]}\nonumber
     \; .
 \end{eqnarray}
 In the thermodynamic limit the free energy is then deduced from the saddle point with respect to $h$ and $q_o$:
 \begin{eqnarray}
    -\beta F_J[\beta,l]=-\beta N f_J[\beta,\, \hat{h} \!, \, \hat{q}_o,\{v_i\},l]   
 \end{eqnarray}
 where $\hat{h}$ and $\hat{q}_o$ are determined by 
 \begin{eqnarray}
 \partial_h f_J\big[\beta,h,q_o,\{v_i\},l\big]\Bigr|_{\substack{h=\hat{h}\\q_o=\hat{q}_o}}=0 \quad \text{and} \quad \partial_{q_o} f_J\big[\beta,h,q_o,\{v_i\},l\big]\Bigr|_{\substack{h=\hat{h}\\q_o=\hat{q}_o}}=0 \; .
 \end{eqnarray}
 
In part of our analysis, we will choose $\{v_i\}$ to be a metastable TAP state at a given temperature $T'$ and it will be designated as the reference state $\{m_i^{\sigma}\}$ while the system described by the spins $\{s_i\}$ will be referred to as the constrained system, and the free energy $-\beta F_J[\beta,q_o,\{v_i\}]$ will be called the constrained free energy. For the moment, we keep $\{v_i\}$ generic.

 \subsection{Taylor expansion of the free energy}
 \label{subsec: Taylor dev}
 
 Following the approach pioneered by Georges and Yedidia \cite{Biroli1999,Georges1991}  we perform a Taylor expansion of the constrained free energy around $\beta=0$ up to second order in $\beta$. Concretely, the series reads
 \begin{eqnarray}
   -\beta F_J[\beta,q_o,\{v_i\},l] & = & \sum_{k=0}^{+\infty}\frac{\beta^k}{k!}\partial^k_\beta (-\beta F_J)\Bigr|_{\beta=0}\\
   & = & -\beta F_J\Bigr|_{\beta=0} +\beta \partial_\beta (-\beta F_J)\Bigr|_{\beta=0} +\frac{\beta^2}{2} \partial^2_\beta (-\beta F_J)\Bigr|_{\beta=0}+O(\beta^3)
   \; . \nonumber
 \end{eqnarray}
Throughout the calculation we will use the notation
\begin{eqnarray}
    \langle \quad.\quad\rangle &=& 
    {\rm Tr}_{\{s_i\}}\Big[\quad.\quad e^{-\beta H_J[\{ s_i \}]-\frac{\lambda}{2}\sum_i (s_i^2-l)-h\sum_i(s_i v_i-q_o)}\Big]/Z_J[\beta,q_o,\{v_i\},l] 
    \; , \\
    \langle \quad.\quad\rangle\Bigr|_{\beta=0}
    &=& 
    {\rm Tr}_{\{s_i\}}\Big[\quad.\quad e^{-\frac{\lambda}{2}\sum_i (s_i^2-l)-h\sum_i(s_i v_i-q_o)}\Big]/Z_J[\beta=0,q_o,\{v_i\},l]
    \; .
\end{eqnarray}
Let us now compute the first terms in the series.

\vspace{0.2cm}

\noindent
\textit{$O^{th}$ order in $\beta$}.
The first term is simply given by the trace over the Gaussian weight
\begin{equation}
    -\beta F_J\Bigr|_{\beta=0}= \ln \Bigg\{  {\rm Tr}_{\{s_i\}}\Big[e^{-\frac{\lambda}{2}\sum_i (s_i^2-l)-h\sum_i(s_i v_i-q_o)}\Big]  \Bigg\}
    \; .
\end{equation}
After integrating over all spin configurations $\{s_i\}$ the previous expression yields (up to a constant)
\begin{eqnarray}
 \label{eq: 0th order TAP}
    -\frac{1}{N}\beta F_J\Bigr|_{\beta=0} & = & \frac{l\lambda}{2}-\frac{\ln\lambda}{2}+h \hspace{0.01cm} q_o+ \frac{h^2}{2\lambda} \frac{1}{N}\sum_i v_i^2 
    \; .
\end{eqnarray}
One can note that at this order
 \begin{equation}
     \langle s_i\rangle\Bigr|_{\beta=0}=
     - \frac{1}{h} \frac{\partial(-\beta F_J|_{\beta=0})}{\partial v_i} 
     =
     \frac{-h v_i}{\lambda}
     \qquad \text{and} \qquad  
     \Big\langle \big(s_i-\langle s_i\rangle\big)^2\Big\rangle\Bigr|_{\beta=0}
     =
     \frac{1}{h^2} \frac{\partial^2(-\beta F_J|_{\beta=0})}{\partial (v_i)^2}
     =
     \frac{1}{\lambda}
     \; .
 \end{equation}
Introducing
\begin{equation}
    q_{v} = \frac{1}{N}\sum_i v_i^2 
    \; 
    \end{equation}
    and using the condition of vanishing variation of the free energy with respect to the Lagrange 
    multipliers: 
    \begin{equation}
        \partial_\lambda \big(-\beta F_J\big)\Bigr|_{\beta=0}=0 =l-\frac{1}{\lambda}- 
        \frac{h^2}{\lambda^2} \, q_v
   \end{equation}   
and
    \begin{equation}
    \partial_h \big(-\beta F_J\big)\Bigr|_{\beta=0}=0 = q_o + \frac{h}{\lambda} \, 
    q_v
    \; ,
\end{equation}
one eliminates $h$ and $\lambda$ to obtain a concise expression for the free energy
\begin{eqnarray}
 \label{eq: 0th order TAP}
    -\beta F_J\Bigr|_{\beta=0} & = &
    \frac{N}{2} 
    \ln\Big(l-\frac{q_o^2}{q_v}\Big)
    \; ,
\end{eqnarray}
and the mean values
 \begin{equation}
 \label{eq: mean high T}
     \langle s_i\rangle\Bigr|_{\beta=0}
     =\frac{q_o}{q_v} v_i \qquad \text{and} \qquad  
     \langle s_i^2\rangle\Bigr|_{\beta=0}
     =l-\frac{q_o^2}{q_v}+\frac{q_o^2}{q_v^2} \, v_i^2
     \; .
 \end{equation}
At this order, the free energy is just the entropy of non-interacting spins lying on the sphere with radius $lN$ and magnetisations $\{(q_o/q_v) v_i\}$. The 
last expression in Eq.~(\ref{eq: mean high T}) implies that the global spherical constraint is preserved.
 
 \vspace{0.2cm}
 \noindent
\textit{$1^{st}$ order in $\beta$:}
The derivation with respect to $\beta$ 
that yields the first order contribution in $\beta$ reads
\begin{eqnarray}
 -\beta \frac{\partial (\beta F_J)}{\partial \beta}\Bigr|_{\beta=0} & = &-\beta \Big\langle H_J[\{s_i\}]+\partial_\beta h\sum_i(s_i v_i-q_o)+\frac{\partial_\beta \lambda}{2}\sum_i({s_i}^2-l)\Big\rangle\Bigr|_{\beta=0}
 \; .
 \end{eqnarray}
 The last two terms vanish as the Lagrangian constraints are verified on average. Taking the limit $\beta=0$ all spins are decoupled, the averages can be explicitly computed, and one finds 
 \begin{eqnarray}
 \label{eq: 1st order TAP}
  -\beta \frac{\partial (\beta F_J)}{\partial \beta}\Bigr|_{\beta=0} & = &  -\beta \Bigg\{ \Big(\frac{q_o}{q_v}\Big)^{p_1} H_{p_1} [\{ v_i \}] + \Big(\frac{q_o}{q_v}\Big)^{p_2} H_{p_2} [\{ v_i \}] \Bigg\}
  \; .
\end{eqnarray}
 Taking $v_i=\langle s_i\rangle$, $q_v=q_o$,  and combining the $0^{th}$ order with the $1^{st}$ order the standard mean field 
 result, in which no overlap constraint is imposed,  is retrieved.

\vspace{0.25cm}
\noindent
\textit{$2^{nd}$ order in $\beta$}.
The second order correction yields the Onsager reaction 
term.
From now on we will consider, for simplicity, the usual case
in which the spherical constraint is set to $l=1$. We will thus drop the dependence of the constrained free energy in $l$ and write it  $-\beta F_J[\beta,q_o,\{v_i\}]$.
The second derivative of the constrained free energy with respect to $\beta$ yields
\begin{eqnarray}
\label{eq:2nd order free energy term(1)}
      \frac{\partial^2 (-\beta F_J)}{\partial \beta^2}& = &\Bigg\langle \Big\{H_J[\{ s_i\}]+\frac{1}{2}\partial_\beta \lambda \sum_i(s_i^2-1)+\partial_\beta h\sum_i(s_i v_i-q_o)\Big\}^2\Bigg\rangle-\Big\langle H_J[\{s_i\}]\Big\rangle^2
            \; .
\end{eqnarray}
 At this point the first order correction in $\beta$ of $h$ and $\lambda$ have to be 
 computed. To do so one can use the Maxwell relations and after some manipulations write: 
\begin{eqnarray}
\label{eq: Maxwell}
       \partial_\beta h\Bigr|_{\beta=0} & = &\frac{1}{N}\partial_\beta\Big[ \partial_{q_o}
       \big(-\beta F_J\big)\Big]\Bigr|_{\beta=0}= \frac{1}{N}\partial_{q_o}\Big[ \partial_\beta\big(-\beta F_J\big)\Big]\Bigr|_{\beta=0} = \frac{1}{N}\partial_{q_o} \Big\langle - H_J[\{ s_i \}] \Big\rangle \Bigr|_{\beta=0}\nonumber\\
       & = & -\frac{1}{N}\partial_{q_o}\Bigg\{ \Big(\frac{q_o}{q_v}\Big)^{p_1} H_{p_1} [\{ v_i \}] + \Big(\frac{q_o}{q_v}\Big)^{p_2} H_{p_2} [\{ v_i\}] \Bigg\}\nonumber\\ 
       &=&- \frac{1}{N} \Bigg\{ p_1 \Big(\frac{{q_o}^{p_1-1}}{{q_v}^{p_1}}\Big) H_{p_1} [\{ v_i \}] + p_2 \Big(\frac{{q_o}^{p_2-1}}{{q_v}^{p_2}}\Big) H_{p_2} [\{ v_i \}] \Bigg\}
       \; ,\\
       \partial_\beta \lambda\Bigr|_{\beta=0} & = &\frac{2}{N}\partial_\beta \Big[\partial_{l}\big(-\beta F_J\big)\Big]\Bigr|_{\beta=0} = \frac{2}{N}\partial_l \Big[\partial_{\beta}\big(-\beta F_J\big)\Big]\Bigr|_{\beta=0} = \frac{2}{N}\partial_{l} \Big\langle - H_J[\{s_i\}] \Big\rangle \Bigr|_{\beta=0} \nonumber\\
       &=& \frac{2}{N}\partial_l \Bigg\{ \Big(\frac{q_o}{q_v}\Big)^{p_1} H_{p_1} [\{ v_i \}] + \Big(\frac{q_o}{q_v}\Big)^{p_2} H_{p_2} [\{ v_i \}] \Bigg\}= 0
       \; .
\end{eqnarray}
The last identity allows us to simplify the second derivative of the free energy that becomes
\begin{eqnarray}
\label{eq:2nd order free energy term}
      \frac{\partial^2 (-\beta F_J)}{\partial \beta^2}\Bigr|_{\beta=0}&  = & \Big\langle (H_J[\{s_i\}])^2 \Big\rangle\Bigr|_{\beta=0}-\Big\langle H_J[\{s_i\}]\Big\rangle\Bigr|_{\beta=0}^2+2\partial_\beta h\Big\langle H_J[\{s_i\}]\sum_i(s_i v_i-q_o)\Big\rangle\Bigr|_{\beta=0}\nonumber\\
      &&+(\partial_\beta h)^2\Bigg\langle\Big[\sum_i(s_i v_i-q_o)\Big]^2\Bigg\rangle\Biggr|_{\beta=0}
      \; .
\end{eqnarray}
 In order to keep the next calculations comprehensible 
 we will introduce a compact notation and rewrite 
 $H_J [\{s_i\}]$ as follows
 \begin{eqnarray}
     H_J[\{ s_i \}]& = &-\sum_{i_1\neq\dots\neq i_{p_1}}\frac{J_{i_1\dots i_{p_1}}}{p_1 !} s_{i_1}\dots s_{i_{p_1}}-\sum_{i_1\neq\dots\neq i_{p_2}}\frac{J_{i_1\dots i_{p_2}}}{p_2 !}s_{i_1}\dots s_{i_{p_2}}
     \nonumber\\
     & = &-\sum_{\{i_{p_1}\}}J_{\{i_{p_1}\}} S_{\{i_{p_1}\}}-\sum_{\{i_{p_2}\}}J_{\{i_{p_2}\}}S_{\{i_{p_2}\}}
 \end{eqnarray}
 with
 \begin{eqnarray}
 \begin{array}{rclrclrcl}
       \{i_{p_1}\} &\equiv& i_1 \neq \dots \neq i_{p_1}
       \;,  
        & \qquad \{k,i_{2},i_{p_1}\} &\equiv& k\neq i_2 \neq \dots \neq i_{p_1} \; ,
        & \qquad J_{\{i_{p_1}\}} &=&\frac{J_{i_1\dots i_{p_1}}}{p_1 !} \; ,
        \\
        S_{\{i_{p_1}\}} &=& s_{i_1}\dots s_{i_{p_1}}
        \; , & \qquad  
        S_{\{i_{2},i_{p_1}\}} &=& s_{i_2}\dots s_{i_{p_1}} \; ,
        & \qquad \Delta S_{\{i_{p_1}\}} &=& \langle{S_{\{i_{p_1}\}}}^2\rangle-{\langle S_{\{i_{p_1}\}}\rangle}^2 \; ,
        \\
        V_{\{i_{2},i_{p_1}\}} &=& v_{i_2} \dots v_{i_{p_1}} \; . 
        & &&
           \end{array}
 \end{eqnarray}
We now proceed to evaluate each term in Eq.~(\ref{eq:2nd order free energy term}) separately; the details of the calculations can be found in App.~\ref{sec:appA}. To begin with we focus on the extensive contribution of the variance of $H_J[\{ s_i \}]$,
\begin{eqnarray}
 && 
    \left.
    \left[
    \Big\langle (H_J[\{s_i\}])^2 \Big\rangle-\Big\langle H_J[\{s_i\}]\Big\rangle^2 
    \right]
    \right|_{\beta=0} 
     =  \frac{N J_{p_1}^2}{2}\Big[1-\Big(\frac{q_o^2}{q_v}\Big)^{p_1}\Big] +\frac{N J_{p_2}^2}{2}\Big[1-\Big(\frac{q_o^2}{q_v}\Big)^{p_2}\Big]
    \nonumber \\
    & &
    \qquad\qquad
    -\frac{N J_{p_1}^2}{2}\Big(1-\frac{q_o^2}{q_v}\Big)p_1\Big(\frac{q_o^2}{q_v}\Big)^{p_1-1}  -\frac{N J_{p_2}^2}{2}\Big(1-\frac{q_o^2}{q_v}\Big)p_2\Big(\frac{q_o^2}{q_v}\Big)^{p_2-1}  \nonumber\\
    & &
    \qquad\qquad
    +\Big(1-\frac{q_o^2}{q_v}\Big)\sum_k \left[\sum_{ \{i_2,i_{p_1}\}} J_{\{k,i_{2},i_{p_1}\}} \Big(\frac{{q_o}}{q_v}\Big)^{p_1-1} V_{\{i_{2},i_{p_1}\}}+ \sum_{ \{j_2,j_{p_2}\}} J_{\{k,j_{2},j_{p_2}\}} \Big(\frac{{q_o}}{q_v}\Big)^{p_2-1} V_{\{j_{2},j_{p_2}\}} \right]^2
    \; . 
    \label{eq: 2nd order TAP(1)}
\end{eqnarray}
The remaining terms in Eq.~(\ref{eq:2nd order free energy term}) yield 
\begin{eqnarray}
 \label{eq: 2nd order TAP(2)}
 \left.
(\partial_\beta h)^2 \Bigg\langle\Big[\sum_i(s_i v_i-q_o)\Big]^2\Bigg\rangle \right|_{\beta=0} & = & N \left(\partial_\beta h\Bigr|_{\beta=0}\right)^2\Big(1-\frac{q_o^2}{q_v}\Big)q_v
\end{eqnarray}
and
 \begin{eqnarray}
  \label{eq: 2nd order TAP(3)}
  \left.
      (\partial_\beta h) \Big\langle H_J[\{s_i\}]\sum_i(s_i v_i-q_o)\Big\rangle
      \right|_{\beta=0}
      & = & {{\partial_\beta h}\Bigr|_{\beta=0}}\hspace{0.1cm}p_1 \Big(1-\frac{q_o^2}{q_v}\Big) \Big(\frac{q_o}{q_v}\Big)^{p_1-1}H_{p_1}[\{ v_i \}] \nonumber\\
      &  &  + {{\partial_\beta h}\Bigr|_{\beta=0}}\hspace{0.1cm}p_2 \Big(1-\frac{q_o^2}{q_v}\Big) \Big(\frac{q_o}{q_v}\Big)^{p_2-1}H_{p_2}[\{ v_i \}] \;.
 \end{eqnarray}
where Eq.~(\ref{eq: Maxwell}) can be used to replace ${{\partial_\beta h}\Bigr|_{\beta=0}}$. Finally, gathering all three orders of the Taylor expansion, Eqs.~(\ref{eq: 0th order TAP}), (\ref{eq: 1st order TAP}), (\ref{eq: 2nd order TAP(1)}), (\ref{eq: 2nd order TAP(2)}), (\ref{eq: 2nd order TAP(3)}), the constrained free energy becomes
\begin{eqnarray}
\label{eq:free energy}
       -\beta F_J[\beta,q_o,\{v_i\}]&=&\frac{N}{2}\ln\Big(1-\frac{q_o^2}{q_v}\Big)-\beta \Bigg\{ \Big(\frac{q_o}{q_v}\Big)^{p_1} H_{p_1} [\{ v_i \}] + \Big(\frac{q_o}{q_v}\Big)^{p_2} H_{p_2} [\{ v_i \}] \Bigg\}\nonumber\\
       & &+\frac{N \beta^2 J_{p_1}^2}{4}\Big[1-p_1\Big(\frac{q_o^2}{q_v}\Big)^{p_1-1}+(p_1-1)\Big(\frac{q_o^2}{q_v}\Big)^{p_1}\Big]  \nonumber\\
       & &+\frac{N \beta^2 J_{p_2}^2}{4}\Big[1-p_2\Big(\frac{q_o^2}{q_v}\Big)^{p_2-1}+(p_2-1)\Big(\frac{q_o^2}{q_v}\Big)^{p_2}\Big]  \nonumber\\
       & &+\frac{\beta^2}{2}\Big(1-\frac{q_o^2}{q_v}\Big)\sum_k \Big[\sum_{ \{i_2,i_{p_1}\}} J_{\{k,i_{2},i_{p_1}\}} \Big(\frac{{q_o}}{q_v}\Big)^{p_1-1} V_{\{i_{2},i_{p_1}\}} \nonumber\\
       & & \hspace{2.8cm}+ \sum_{ \{j_2,j_{p_2}\}} J_{\{k,j_{2},j_{p_2}\}} \Big(\frac{{q_o}}{q_v}\Big)^{p_2-1} V_{\{j_{2},j_{p_2}\}} \Big]^2 \nonumber\\
       &  &-\frac{\beta^2}{2 q_v N} \Big(1-\frac{q_o^2}{q_v}\Big)\Bigg\{ p_1 \Big(\frac{q_o}{q_v}\Big)^{p_1-1} H_{p_1} [\{ v_i \}] + p_2 \Big(\frac{q_o}{q_v}\Big)^{p_2-1} H_{p_2} [\{ v_i \}] \Bigg\}^2 +O(\beta^3) \; .
\end{eqnarray}
One can rewrite this expression under the form
\begin{eqnarray}
\label{eq:free energy(1)}
       -\beta F_J[\beta,q_o,\{v_i\}]&=&
       -\beta F_{\rm TAP}\Big[\beta,\Big\{\frac{q_o}{q_v} v_i\Big\}\Big]\nonumber\\
       & &+\frac{\beta^2}{2}\Big(1-\frac{q_o^2}{q_v}\Big)\sum_k \Big[\sum_{ \{i_2,i_{p_1}\}} J_{\{k,i_{2},i_{p_1}\}}\Big(\frac{{q_o}}{q_v}\Big)^{p_1-1} V_{\{i_{2},i_{p_1}\}} \nonumber\\
       & & \hspace{2.8cm}+ \sum_{ \{j_2,j_{p_2}\}} J_{\{k,j_{2},j_{p_2}\}} \Big(\frac{{q_o}}{q_v}\Big)^{p_2-1} V_{\{j_{2},j_{p_2}\}} \Big]^2 \nonumber\\
       &  &-\frac{\beta^2}{2 q_v N} \Big(1-\frac{q_o^2}{q_v}\Big)\Bigg\{ p_1 \Big(\frac{q_o}{q_v}\Big)^{p_1-1} H_{p_1} [\{ v_i \}] + p_2 \Big(\frac{q_o}{q_v}\Big)^{p_2-1} H_{p_2} [\{ v_i \}] \Bigg\}^2 +O(\beta^3)
\end{eqnarray}
where $-\beta F_{\rm TAP}$ is the unconstrained TAP free energy for the mixed model 
\begin{eqnarray}
-\beta F_{\rm TAP}[\beta,\{m_i\}]&=&\frac{N}{2}\ln(1-q)-\beta  \Big( H_{p_1} [\{ m_i \}] +  H_{p_2} [\{ m_i \}] \Big)\nonumber\\
       & &+\frac{N \beta^2 J_{p_1}^2}{4}\Big[1-p_1 q^{p_1-1}+(p_1-1) q^{p_1}\Big]  
       +\frac{N \beta^2 J_{p_2}^2}{4}\Big[1-p_2 q^{p_2-1}+(p_2-1) q^{p_2}\Big] 
       \label{eq:fTAP}
\end{eqnarray}
with 
\begin{equation}
 m_i=\frac{q_o}{q_v}v_i \quad \text{and} \quad   q=\frac{1}{N}\sum_i m_i^2
    \; .
    \label{eq:q-def}
\end{equation}
(In~\cite{Annibale2004} this same $F_{\rm TAP}$ appears and it is presented as the result of a perturbative expansion in which one of the two $p$
Hamiltonian's is treated as a perturbation with respect to the other one. In~\cite{Rizzo2006} the TAP free energy is considered to be exact to order $N$ and it describes exactly the statics of the model.)

Fixing the reference $\{v_i\}$ and the temperature $T$ for the system one can note that the constrained free energy only depends on the overlap $q_o$ and not on an extensive number of parameters like is the case in the usual TAP free energy. Here, however, we will have to keep track of the choice of the reference state. We finally emphasise that this derivation differs from the usual TAP calculation as the constrained free energy is determined only up to $O(\beta^3)$ correction terms that we cannot ensure are subleading in $N$. Indeed we have exchanged the local fields $\{h_i\}$ of the TAP method (see App.~\ref{sec: app TAP free energy}) with a global field $h$. Thus the cavity method arguments (perturbing the system with one incremented spin) or the diagrammatic expansion in Ref.~\cite{Crisanti1995} do not apply here as the previous local feature arising with the fields $\{h_i\}$ is lost.



\subsection{A particular case: the pure $p$-spin model}

In the case of the pure $p$-spin model, with a single term in the Hamiltonian $H_J[\{s_i\}]=H_p[\{s_i\}]$, the constrained 
free energy simplifies drastically.  Moreover, taking $\{v_i=m_i^\sigma\}$ a metastable TAP state at temperature $T'$,  
the stationary points of the constrained free energy are of two kinds: either the system keeps a non-vanishing overlap with the reference state, $q_o\neq 0$, 
or it becomes paramagnetic, $q_o=0$.  In the dynamic interpretation of this approach, the former situation is linked to the possibility of following the initial state in a, say, low temperature quench while the latter corresponds to escaping the non-trivial TAP state towards the disordered paramagnetic phase.

To begin with, one can note that if the reference state $\{m_i^\sigma\}$ is metastable at a temperature $T'=1/\beta'$,
it should verify the TAP equations
\begin{eqnarray}
    \frac{m_k^\sigma}{1-q_\sigma}&=&\beta'\sum_{\{i_2,i_{p}\}} J_{k,\{i_{2},i_{p}\}} M_{\{i_{2},i_{p}\}}
    -{\beta'^2 J^2}(1-q_\sigma)\frac{p(p-1)}{2} {q_\sigma}^{p-2} m_k^\sigma
    \qquad \text{with} \qquad k=1,\dots p
    \; , 
\end{eqnarray}
 $M_{i_2,i_p}= m_{i_2}^\sigma \dots m_{i_p}^\sigma$ and $q_\sigma=N^{-1}\sum_i (m_i^\sigma)^2$,
obtained as a stationary point condition 
on $F_{\rm TAP}$ in Eq.~(\ref{eq:fTAP}) with $J_2=0$, $J_1=J$ and $p_1=p$. This equation leads straightforwardly to
\begin{eqnarray}
    \frac{Nq_\sigma}{1-q_\sigma}&=&-\beta' p H_{p} [\{ m_i^\sigma \}]-{\beta'^2 J^2} N (1-q_\sigma)\frac{p(p-1)}{2} q_\sigma^{p-1} 
    \; .
\end{eqnarray}
Again, a lengthy calculation shows that the last two terms in Eq.~(\ref{eq:free energy}) cancel out and the constrained free energy becomes
\begin{eqnarray}
\label{eq: free energy pure p spin}
       -\beta F_J [\beta,q_o,\{m_i^{\sigma}\}]&=&\frac{N}{2}\ln(1-\frac{q_o^2}{q_\sigma})-\beta  \Big(\frac{q_o}{q_\sigma}\Big)^{p} H_p[\{ m_i^\sigma \}] +\frac{N \beta^2 {J^2}}{4}\Big[1-p\Big(\frac{q_o^2}{q_\sigma}\Big)^{p-1}+(p-1)\Big(\frac{q_o^2}{q_\sigma}\Big)^{p}\Big]+O(\beta^3)\nonumber\\
       &=&-\beta F_{\rm TAP}\Big[\beta,\Big\{\frac{q_o}{q_\sigma} m_i^{\sigma}\Big\}\Big]+O(\beta^3)\; ,
\end{eqnarray}
that is to say,  the TAP free energy for a $p$-spin model with local magnetisations and overlap
\begin{equation}
    m_i=\Big(\frac{q_o}{q_\sigma}\Big) m_i^\sigma \qquad \text{and} \qquad q=\frac{1}{N}\sum_i m_i^2=\frac{q_o^2}{q_\sigma}
    \; ,
\end{equation}
respectively.
 In fact, as detailed in App.~\ref{sec:app_link_TAP_constrained}, the constrained free energy $-\beta F_J [\beta,q_o,\{m_i^{\sigma}\}]$ is strictly equal to the TAP free energy $-\beta F_{\rm TAP}\Big[\beta,\Big\{\frac{q_o}{q_\sigma} m_i^{\sigma}\Big\}\Big]$ in this case. In other words the $O(\beta^3)$ terms and higher order ones vanish in the $N \rightarrow +\infty$ limit, Eq.~(\ref{eq: free energy pure p spin}) is then exact and not approximated. 
 
As previewed at the beginning of this section, the solutions minimising the free energy are such that
\begin{eqnarray}
    &&\partial_{q_o}(-\beta F_J)=\frac{2q_o}{q_\sigma}\partial_{\frac{q_o^2}{q_\sigma}}(-\beta F_J)=0
    \qquad \Rightarrow \qquad
    \left\{
    \begin{array}{l}
     \displaystyle{
     \frac{p(p-1)}{2}z^2+p \epsilon z+1
     = 0 
     } \; , \\
     q_o=0 \; , 
     \end{array}
     \right.
\end{eqnarray}
with
\begin{eqnarray}
    z= \frac{J}{T} \Big(1-\frac{q_o^2}{q_\sigma}\Big)\Big(\frac{q_o^2}{q_\sigma}\Big)^{\frac{p}{2}-1} \qquad \text{and} \qquad \varepsilon =\frac{1}{N J {q_\sigma}^{\frac{p}{2}}}H_J[\{ m_i^\sigma \}]
    \; .
\end{eqnarray}

The interpretation of these solutions, in dynamical terms, is the following.
On the one hand the $q_o\neq 0$ solution corresponds to the system -after the quench- staying in the same TAP metastable state up to the rescaling $ m_i=({q_o}/{q_\sigma}) m_i^\sigma$. On the other hand, with the $q_o=0$ solution one recovers the free energy of the paramagnetic state, it corresponds to a quench to high temperature where the first solution is not available anymore (i.e. the initial TAP state becomes unstable). These conclusions have already been drawn in previous papers, see e.g. Ref.~\cite{Barrat}, using different methods and 
focusing on the dynamics with an initial temperature $T' \in [T_s;T_d]$.

We conclude that the constrained free energy density does not provide further information about the behaviour of the pure model, compared to what had been derived from the unconstrained one.


\section{Application to the mixed $p$-spin model}
\label{sec:Application to the mixed $p$-spin model}

In this section we apply the constrained free energy density to the analysis of the mixed model.

\subsection{Simplification of the free energy and stability condition}

Contrary to the pure $p$-spin model the last two terms of the constrained 
free energy in Eq.~(\ref{eq:free energy(1)}) do not cancel out and can be seen as being at the origin of
some of the peculiar features of these models. 

For the following analysis we will consider the reference state $\{v_i=m_i^\sigma\}$ to be a metastable TAP state at a given temperature $T'=1/\beta'$. Under this choice, 
the constrained free energy can be simplified to -for details, see App.~\ref{sec:appB}-
\begin{equation}
\label{eq:free energy mixed p-spin}
          -\beta F_J [\beta,q_o,\{m_i^{\sigma}\}] =-\beta F_{\rm TAP}\Big[\beta,\Big\{\frac{q_o}{q_\sigma} m_i^{\sigma}\Big\}\Big]+\frac{\beta^2}{2}\Big(1-\frac{q_o^2}{q_\sigma}\Big)\Bigg[\Big(\frac{{q_o}}{q_\sigma}\Big)^{p_2-1}-\Big(\frac{{q_o}}{q_\sigma}\Big)^{p_1-1}\Bigg]^2\Delta H_{p_2}[\{ m_i^\sigma \}]+O(\beta^3)
\end{equation}
where we used the form of $F_{\rm TAP}$ in Eq.~(\ref{eq:fTAP}) and
\begin{equation}
     \Delta H_{p_2}[\{ m_i^\sigma \}]=  \sum_k \Big(\partial_{m_k^\sigma} H_{p_2} [\{ m_i^\sigma \}] \Big)^2-\frac{{p_2}^2}{N}\Big(H_{p_2} [\{ m_i^\sigma \}]\Big)^2=\left\|{\nabla H_{p_2}}\right\|^2-\frac{{p_2}^2}{N}
     \Big( H_{p_2} [\{ m_i^\sigma \}]\Big)^2 \; .
\end{equation}
The extra term $\Delta H_{p_2}[\{ m_i^\sigma \}]$, depending on the reference $\{m_i^\sigma\}$, cannot be rewritten using the usual variables $q_\sigma, $ $H_{p_1} [\{ m_i^\sigma \}]$ and $H_{p_2} [\{ m_i^\sigma \}]$. It is a direct consequence of the Hamiltonian and its non homogeneity. More practically this can be seen with the terms
\begin{eqnarray}
        \sum_{ \{i_2,i_{p_1}\}} J_{\{k,i_{2},i_{p_1}\}}\Big(\frac{{q_o}}{q_v}\Big)^{p_1-1} V_{\{i_{2},i_{p_1}\}} + \sum_{ \{j_2,j_{p_2}\}} J_{\{k,j_{2},j_{p_2}\}} \Big(\frac{{q_o}}{q_v}\Big)^{p_2-1} V_{\{j_{2},j_{p_2}\}} \; 
\end{eqnarray}
appearing in Eq.~(\ref{eq:free energy(1)}).
The dependence in $({q_o}/{q_v})^{p_1-1}$ and $({q_o}/{q_v})^{p_2-1}$ prevent us from simplifications using the TAP equations. Indeed, for simplifications one would rather need  terms of the form -for details, see App.~\ref{sec:appB}-
\begin{eqnarray}
   \sum_{ \{i_2,i_{p_1}\}} J_{\{k,i_{2},i_{p_1}\}} V_{\{i_{2},i_{p_1}\}} + \sum_{ \{j_2,j_{p_2}\}} J_{\{k,j_{2},j_{p_2}\}}  V_{\{j_{2},j_{p_2}\}} \; .  
\end{eqnarray}

As detailed in App.~\ref{sec:app_link_TAP_constrained}, if we impose $q_o=q_\sigma$ the expression for the constrained free energy~(\ref{eq:free energy mixed p-spin}) becomes exact at any order in $\beta$ in the $N\rightarrow +\infty$ limit. Indeed, the $O(\beta^3)$ term yields sub-extensive contributions to the free energy.

For the general case, up to $O(\beta^3)$ corrections, any metastable state should minimise the free energy with respect to $q_o$ such that
\begin{eqnarray}
\label{eq: stability}
       \partial_{q_o}\Bigg\{  -\beta F_{\rm TAP}\Big[\beta,\Big\{\frac{q_o}{q_\sigma} m_i^{\sigma}\Big\}\Big] \Bigg\}+  \partial_{q_o}\Bigg\{\Big(1-\frac{q_o^2}{q_\sigma}\Big)\Bigg[\Big(\frac{{q_o}}{q_\sigma}\Big)^{p_2-1}-\Big(\frac{{q_o}}{q_\sigma}\Big)^{p_1-1}\Bigg]^2\Delta H_{p_2}[\{ m_i^\sigma \}]\Bigg\}=0
\end{eqnarray}
Looking at the case $\beta=\beta'$ one can check that $q_o=q_\sigma$ is a stationary point of the constrained free energy yielding simply $F_J [\beta',q_\sigma,\{m_i^{\sigma}\}]=F_{\rm TAP}[\beta',\{m_i^{\sigma}\}]$.
We recover here the expected result that the reference state $\{m_i^{\sigma}\}$ is metastable for $\beta=\beta'$. However, the general situation (for any $\beta,\beta'$) is non-trivial. In fact, taking the variation of the constrained free energy with respect to $q_o$ yields the stationary conditions
\begin{eqnarray}
\label{eq:stability}
\partial_{q_o}(-\beta F_J) &=& 0\\
&\Longleftrightarrow &\left\{
    \begin{array}{lll}
        q_o=0 \\
        \\
        0=1+\sum_{\ell=1}^2 \Big\{\frac{p_\ell(p_\ell-1)}{2}z_\ell^2+p \varepsilon_{p_\ell}[\{ m_i^\sigma \}] z_\ell\Big\} \\
        \hspace{0.65cm}+{q_\sigma} \frac{\Delta H_{p_2}[\{ m_i^\sigma \}]}{N}   \Bigg[\displaystyle{\frac{ z_2}{J_{p_2}{q_\sigma}^{\frac{p_2}{2}}}-\frac{ z_1}{J_{p_1}{q_\sigma}^{\frac{p_1}{2}}}}\Bigg]  \Bigg[\frac{ z_2}{J_{p_2}{q_\sigma}^{\frac{p_2}{2}}}\Big(p_2-1-\frac{1}{\frac{q_\sigma}{q_o^2}-1}\Big)-\frac{z_1}{J_{p_1}{q_\sigma}^{\frac{p_1}{2}}}\Big(p_1-1-\frac{1}{\frac{q_\sigma}{q_o^2}-1}\Big)\Bigg]  \\
    \end{array}
    \right.\nonumber
\end{eqnarray}
with
\begin{eqnarray}
    z_\ell= {J_{p_\ell} \beta} \Big(1-\frac{q_o^2}{q_\sigma}\Big)\Big(\frac{q_o^2}{q_\sigma}\Big)^{\frac{p_\ell}{2}-1} \quad \text{and} \quad \varepsilon_{p_\ell} [\{ m_i^\sigma \}] =\frac{1}{N J_{p_\ell} {q_\sigma}^{\frac{p_\ell}{2}}}H_{p_\ell} [\{ m_i^\sigma \}] \quad \qquad \mbox{for} \qquad \ell = 1,2 
    \; . 
\end{eqnarray}

\subsection{Discussion}

If we now focus on a reference $\{v_i = m_i^\sigma\}$ 
that is a metastable TAP state, the constrained free energy $-\beta F_J[\beta,q_o,\{m_i^{\sigma}\}]$  makes the chaos in temperature appear clearly. In particular it shows that different states $\{m_i^{\sigma}\}$ yield different constrained systems. To see these features
we compute the total energy,
\begin{eqnarray}
\label{eq:energy_shift}
    E_J[\beta,q_o,\{m_i^{\sigma}\}] 
    &=&
    -\partial_\beta  \Bigg\{-\beta F_J[\beta,q_o,\{m_i^{\sigma}\}]\Bigg\}\\ 
    &=&E_{\rm TAP}\Big[\beta,\Big\{\frac{q_o}{q_\sigma}m_i^{\sigma}\Big\}\Big]+\beta\Big(1-\frac{q_o^2}{q_\sigma}\Big)\Bigg[\Big(\frac{{q_o}}{q_\sigma}\Big)^{p_2-1}-\Big(\frac{{q_o}}{q_\sigma}\Big)^{p_1-1}\Bigg]^2\Delta H_{p_2}[\{ m_i^\sigma \}] 
    \nonumber
    \; .
\end{eqnarray}
The first term, $E_{\rm TAP}\Big[\beta,\Big\{\frac{q_o}{q_\sigma}m_i^{\sigma}\Big\}\Big]$, 
is the energy that the system would have if it were at temperature $T$
within the same TAP state as the reference was. The sole change in energy would then be given by the state deformation represented by the renormalisation  $m_i^\sigma \rightarrow (q_o/q_\sigma) \, m_i^{\sigma}$.
The information about the 
shift of TAP state from the reference to another one is thus contained in the second term in Eq.~(\ref{eq:energy_shift}),  as it implies $E_J[\beta,q_o,\{m_i^\sigma\}] \neq E_{\rm TAP}\Big[\beta,\Big\{\frac{q_o}{q_\sigma}m_i^{\sigma}\Big\}\Big]$. We note that the constrained system depends  explicitly on the reference state $\{m_i^\sigma\}$ via the value taken by $\Delta H_{p_2}[\{ m_i^\sigma \}]$. Again it is important to point out that, as the mixed $p$-spin glass model has a non homogeneous Hamiltonian, this term cannot be rewritten using the usual equilibrium parameter $q_\sigma$, $H_{p_1}[\{m_i^\sigma\}]$ and $H_{p_2}[\{m_i^\sigma\}]$. In the case of the pure $p$-spin model
$\Delta H_{p_2}[\{ m_i^\sigma \}]=0$ and there is no shift of TAP state.

\begin{figure}[h]
    \centering
    \includegraphics[width=0.7\textwidth]{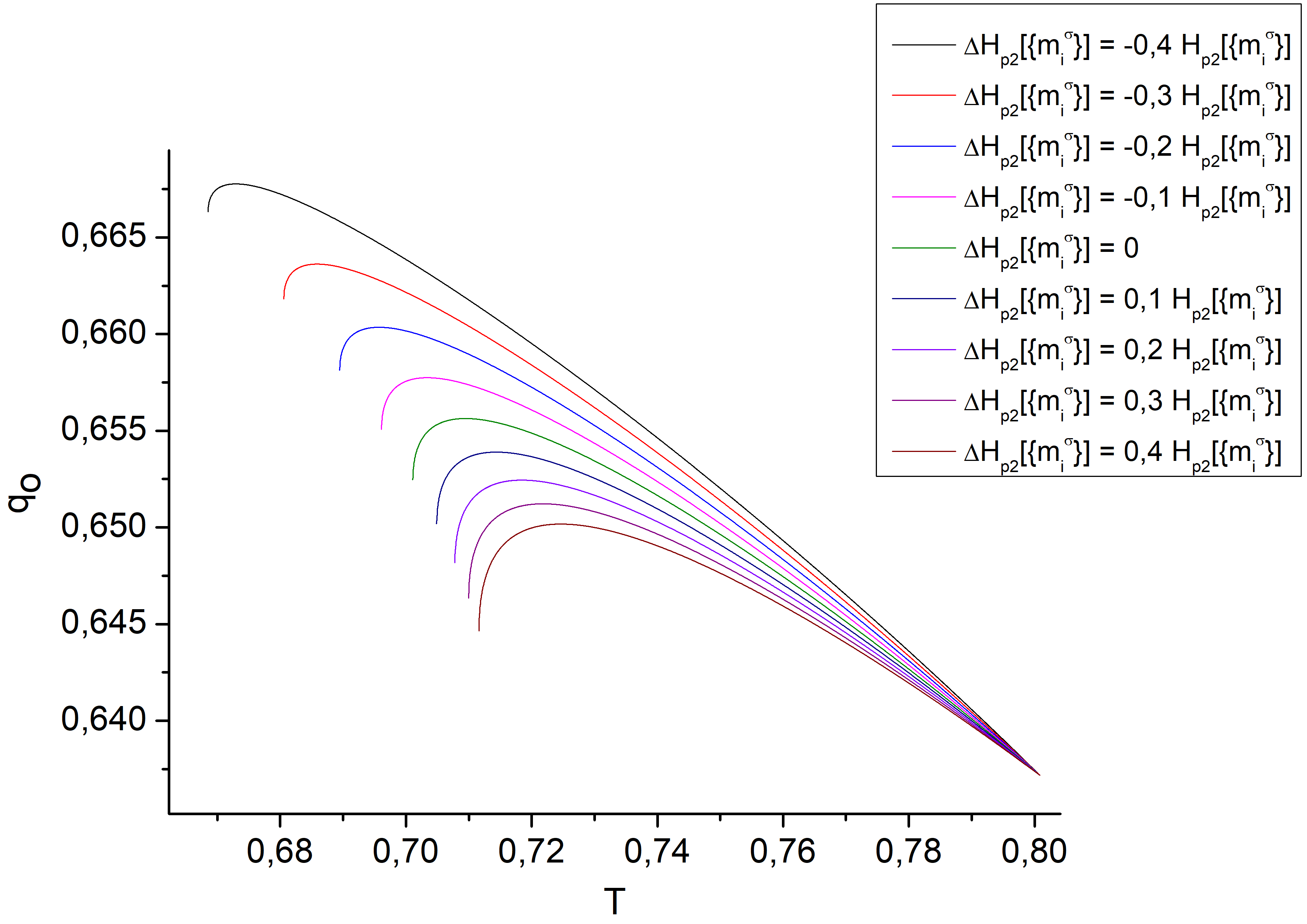}
    \caption{Changing the temperature $T=1/\beta$, we plotted the $q_o$ that minimises the constrained free energy -see Eq.~(\ref{eq:stability}). For the reference $\{m_i^\sigma\}$ we focused on the metastable TAP states at $T'=1/\beta'$ that dominate the Gibbs-Boltzmann distribution $P[\{s_i\}]\propto \exp\big(-\beta' H_J[\{s_i\}]\big)$. We recall that they are fully parametrised by their overlap $q = N^{-1} \sum_i m_i^2$ and the adimensional energy densities $\varepsilon_{p_1}= N^{-1} J_{p_1}^{-1} q^{-p_1/2} H_{p_1}[\{m_i\}]$ 
and $\varepsilon_{p_2}=N^{-1}J_{p_2}^{-1} q^{-p_2/2} H_{p_2}[\{m_i\}]$. However, $\Delta H_{p_2}[\{ m_i^\sigma \}]$ is a priori not fixed and was taken to have the values given in the key.  As parameters we fixed $T'\approx 0.801$, $p_1=3$, $p_2=4$ and $J_{p_1}=J_{p_2}=1$. We took  for $H_{p_1}[\{m_i^\sigma\}]$ and $H_{p_2}[\{m_i^\sigma\}]$ the values given by the equilibrated system at $T'$. }
    \label{fig:TAP_quench}
\end{figure}

Chaos in temperature can also be observed through the minimisation of the constrained free energy. Indeed the value of $q_o$ determined by Eq.~(\ref{eq:stability}) depends on the reference $\{m_i^\sigma\}$ again via the ``parameter" $\Delta H_{p_2}[\{ m_i^\sigma \}]$ -see Fig.~\ref{fig:TAP_quench}. Consequently different TAP states impose different overlaps $q_o$, besides yielding  constrained systems with different energies.

Interpreting this result in dynamic terms, a system initially equilibrated at $T'$ and then quenched to a temperature $T \in [T_{\rm RSB}(T'),T'[$ departs in different metastable states depending on which TAP state $\{m_i^{\sigma}\}$ it was initially laid in. This is different from what happens in the pure $p$-spin model, in which metastable TAP states can be fully followed in temperature.
This chaos in temperature cannot be observed using the FP potential nor using the Schwinger-Dyson equations \cite{Capone2006} (in the context of dynamics). Indeed both procedures average over the constraining system (respectively the initial system), thus making it impossible to keep track of each reference state. This chaotic behavior of the mixed $p$-spin model may also be useful to interpret the strange dynamics observed for quenches to low temperatures ($T<T_{\rm RSB}(T')$)  and  zero temperature dynamics \cite{Folena2019}.

\section{An exact approach for the constrained free energy}
\label{se: an exact approach for the constrained free energy}
In the following section we will present a method to derive exactly the constrained free energy. It will consist in mapping the TAP free energy on the constrained one. Via the high temperature expansion we have already shown  the equivalence between the two free energies when $q_o=q_\sigma$; the constrained system is a TAP state with local magnetisation $\langle s_i\rangle=m_i^\sigma$ in this situation. To generalise this result to any value of $q_o$  we will start by considering the Legendre transforms of the constrained and TAP free energies -see Eqs.~(\ref{eq: free energy def}) and~(\ref{eq: free energy def2}): 

\begin{eqnarray}
        -\beta F_J^\star[\beta,\{h v_i\},\lambda]&=&-\beta F_J [\beta,q_o,\{v_i\},l]-\frac{N \lambda l}{2}-N h q_o\\
        &=&\ln \Bigg\{  {\rm Tr}_{\{s_i\}}\Big[e^{-\beta H_J[\{s_i\}]-\frac{\lambda}{2}\sum_i (s_i^2-l)-h\sum_i(s_i v_i-q_o)}\Big]  \Bigg\}-\frac{N \lambda l}{2}-N h q_o\nonumber\\
        &=&\ln \Bigg\{  {\rm Tr}_{\{s_i\}}\Big[e^{-\beta H_J[\{s_i\}]-\frac{\lambda}{2}\sum_i (s_i^2)-\sum_i (h v_i) s_i}\Big]  \Bigg\} \;,\nonumber
\end{eqnarray}
\begin{eqnarray}
        -\beta F_{\rm TAP}^\star[\beta,\{h_i\},\lambda']&=&-\beta F_{\rm TAP}[\beta,\{m_i\},l]-\frac{N \lambda' l}{2}-\sum_i h_i m_i\\
        &=&\ln \Bigg\{  {\rm Tr}_{\{s_i\}}\Big[e^{-\beta H_{J}[\{s_i\}]-\frac{\lambda'}{2}\sum_i (s_i^2-l)-\sum_ih_i(s_i -m_i)}\Big]  \Bigg\}-\frac{N \lambda' l}{2}-\sum_i h_i m_i \nonumber\\
        &=&\ln \Bigg\{  {\rm Tr}_{\{s_i\}}\Big[e^{-\beta H_J[\{s_i\}]-\frac{\lambda'}{2}\sum_i (s_i^2)-\sum_i h_i s_i}\Big]  \Bigg\} \;.\nonumber
\end{eqnarray}
The first step of our approach is to Taylor expand the two free energies in orders of $\beta$ keeping the Lagrange multipliers $\lambda'$, $\lambda$, $\{h_i\}$ and $h$ constants. In more details we write
\begin{equation}
-\beta F_J^\star[\beta,\{h v_i\},\lambda]  =  \sum_{k=0}^{+\infty}\frac{\beta^k}{k!}\partial^k_\beta \Big(-\beta F_J^\star[\beta,\{h v_i\},\lambda]\Big)\Bigr|_{\beta=0}
\end{equation}
and
\begin{equation}
-\beta F_{\rm TAP}^\star[\beta,\{h_i\},\lambda']  =  \sum_{k=0}^{+\infty}\frac{\beta^k}{k!}\partial^k_\beta \Big(-\beta F_{\rm TAP}^\star[\beta,\{h_i\},\lambda']\Big)\Bigr|_{\beta=0} \; .
\end{equation}
The $0^{\rm th}$ order is a Gaussian integral in both cases, it is almost identical to the $0^{\rm th}$ order expansion in Sec.~\ref{subsec: Taylor dev}. We have  straightforwardly
\begin{eqnarray}
    F_J^\star[\beta=0,\{h v_i\},\lambda]=-\frac{N}{2}\ln(\lambda)+\sum_i \frac{h^2 v_i^2}{2\lambda} \; ,
\end{eqnarray}
\begin{eqnarray}
    F_{\rm TAP}^\star[\beta=0,\{h_i\},\lambda']=-\frac{N}{2}\ln(\lambda')+\sum_i \frac{h_i^2}{2\lambda'} \; .
\end{eqnarray}
For the constrained free energy the following orders are simply functions of $\lambda$ and $\{h v_i\}$, they are of the generic form 
\begin{eqnarray}
         \partial_\beta^k \Big(-\beta F_J^\star[\beta,\{h v_i\},\lambda]\Big)\Bigr|_{\beta=0}&=&\Big\langle A_k [\{s_i\}]\Big\rangle\Bigr|_{\beta=0}+ B_k \Big[\Big\{\langle s_i\rangle\bigr|_{\beta=0}\Big\}\Big]\\
         &=&A'_k \Big[\Big\{\langle s_i\rangle\bigr|_{\beta=0} \; ; \;\langle s_i^2\rangle\bigr|_{\beta=0}\Big\}\Big]+ B_k \Big[\Big\{\langle s_i\rangle\bigr|_{\beta=0}\Big\}\Big] \nonumber
\end{eqnarray}
with 
\begin{eqnarray}
\label{eq:average1}
    \langle s_i\rangle\Bigr|_{\beta=0}
    &=& 
    -\frac{h v_i}{\lambda} \quad \text{and} \quad  \langle s_i^2\rangle\Bigr|_{\beta=0}
    = 
    \frac{1}{\lambda}+\frac{h^2 v_i^2}{\lambda^2} \; .
\end{eqnarray}
The case of the TAP free energy is equivalent, we have
\begin{eqnarray}
         \partial_\beta^k \Big(-\beta F_{\rm TAP}^\star[\beta,\{h_i\},\lambda']\Big)\Bigr|_{\beta=0}&=&\Big\langle A_k [\{s_i\}]\Big\rangle\Bigr|_{\beta=0}+ B_k \Big[\Big\{\langle s_i\rangle\bigr|_{\beta=0}\Big\}\Big]\\
         &=&A'_k \Big[\Big\{\langle s_i\rangle\bigr|_{\beta=0} \; ; \;\langle s_i^2\rangle\bigr|_{\beta=0}\Big\}\Big]+ B_k \Big[\Big\{\langle s_i\rangle\bigr|_{\beta=0}\Big\}\Big]\nonumber
\end{eqnarray}
with 
\begin{eqnarray}
\label{eq:average2}
    \langle s_i\rangle\Bigr|_{\beta=0}
    &=& 
    -\frac{h_i}{\lambda'} \quad \text{and} \quad  \langle s_i^2\rangle\Bigr|_{\beta=0}
    = 
    \frac{1}{\lambda'}+\frac{h_i^2}{\lambda'^2} \; .
\end{eqnarray}
As an example the $1^{\rm st}$ order is
\begin{eqnarray}
      & \partial_\beta \Big(-\beta F_J^\star[\beta,\{h v_i\},\lambda]\Big)\Bigr|_{\beta=0}=-\Big\langle H_J [\{s_i\}]\Big\rangle\Bigr|_{\beta=0}=-H_J\big[\{\frac{h v_i}{\lambda}\}\big] 
\end{eqnarray}
and
\begin{eqnarray}
      & \partial_\beta \Big(-\beta F_{\rm TAP}^\star[\beta=0,\{h_i\},\lambda']\Big)\Bigr|_{\beta}=-\Big\langle H_J [\{s_i\}]\Big\rangle\Bigr|_{\beta=0} =-H_J\big[\{\frac{h_i}{\lambda'}\}\big] \; .
\end{eqnarray}
At this stage it is important to note that the expansion in terms of the functions $A'_k$ and $B_k$ is identical for both Legendre transforms, thus they differ from each other only through their spin averages (\ref{eq:average1}),(\ref{eq:average2}). The next step of our reasoning is to set $\lambda=\lambda'$ and $h_i=h v_i$ for all the local fields, then the Legendre transforms become equal:
\begin{eqnarray}
\label{eq: equality1}
        -\beta F_J^\star[\beta,\{h v_i\},\lambda]=-\beta F_{\rm TAP}^\star[\beta,\{h v_i\},\lambda]\; .
\end{eqnarray}
For the TAP free energy the Taylor expansion is known exactly when we set
\begin{eqnarray}
       h_i&=&\frac{-m_i}{l-q}-\beta \partial_{m_i} H_J [\{m_j\}]-\frac{p_1(p_1-1)\beta^2 J_{p_1}^2}{2}(l-q)q^{p_1-2}m_i-\frac{p_2(p_2-1)\beta^2 J_{p_2}^2}{2}(l-q)q^{p_2-2}m_i \; ,\\
       \lambda&=&\frac{1}{l-q}+\frac{p_1\beta^2 J_{p_1}^2}{2}(l^{p_1-1}-q^{p_1-1})+\frac{p_2\beta^2 J_{p_2}^2}{2}(l^{p_2-1}-q^{p_2-1})
\end{eqnarray}
with $q=\sum_i m_i^2$. Thus we retrieve Eq.~(\ref{eq:free_energy_TAP}) 
\begin{eqnarray}
\label{eq: equality2}
      -\beta F_{\rm TAP}^\star[\beta,\{h_i\},\lambda] &=&
      -\frac{N \lambda l}{2}-\sum_i h_i m_i-\beta F_{\rm TAP}[\beta,\{m_i\},l] \\
     &=&-\frac{N \lambda l}{2}-\sum_i h_i m_i-\frac{NT}{2}\ln(l-q)+ N q^{\frac{p_1}{2}} \, J_{p_1} \, \varepsilon_{p_1} +N q^{\frac{p_2}{2}} \, J_{p_2} \, \varepsilon_{p_2}
    \nonumber\\
    & &  -\frac{N J_{p_1}^2}{4T}\Big[l^{p_1}-p_1 l q^{p_1-1}+(p_1-1)q^{p_1}\Big]-\frac{N J_2^2}{4T}\Big[l^{p_2}-p_2 l q^{p_2-1}+(p_2-1)q^{p_2}\Big] \; . \nonumber
\end{eqnarray}
Combining Eqs.~(\ref{eq: equality1}) and (\ref{eq: equality2}) we can finally write
\begin{eqnarray}
      -\beta F_J^\star[\beta,\{h v_i\},\lambda] =-\beta F_{\rm TAP}^\star[\beta,\{h v_i\},\lambda]=-\beta F_{\rm TAP}[\beta,\{m_i\},l] -\frac{N \lambda l}{2}-h\sum_i v_i m_i
\end{eqnarray}
with the prescription
\begin{eqnarray}
\label{eq:h vs magnetisations}
       h v_i&=&\frac{-m_i}{l-q}-\beta \partial_{m_i} H_J [\{m_j\}]-\frac{p_1(p_1-1)\beta^2 J_{p_1}^2}{2}(l-q)q^{p_1-2}m_i-\frac{p_2(p_2-1)\beta^2 J_{p_2}^2}{2}(l-q)q^{p_2-2}m_i \; ,\\
       \lambda&=&\frac{1}{l-q}+\frac{p_1\beta^2 J_{p_1}^2}{2}(l^{p_1-1}-q^{p_1-1}) \; .
\end{eqnarray}
To satisfy the spherical constraint and the overlap with the reference we write Eqs.~(\ref{eq: constraining the free energy}) and (\ref{eq: constraining the free energy(2)}) as follow
\begin{eqnarray}
       \partial_\lambda \big(-\beta F_J^\star[\beta,\{h v_i\},\lambda]\big)    
       &=&    
       \partial_\lambda \big(-\beta F_{\rm TAP}^\star[\beta,\{h v_i\},\lambda]\big)=\frac{N l}{2} \implies l = 1 \; ,
\\
       \partial_h \big(-\beta F_J^\star[\beta,\{h v_i\},\lambda]\big)
       &=&    
       \sum_i v_i \partial_{hv_i} \big(-\beta F_{\rm TAP}^\star[\beta,\{h v_i\},\lambda]\big)=\sum_i v_i m_i \;\implies  \sum_i v_i m_i = N q_o \; .
\end{eqnarray}
We can also note, as explained in subsection \ref{sec: justification of the approach}, that extremising  the constrained free energy with respect to $q_o$ is equivalent to setting $h=0$. It follows straightforwardly from Eq.~(\ref{eq:h vs magnetisations}) that the state $\{m_i\}$ obtained by minimising the constrained free energy is a metastable TAP state that verifies
\begin{eqnarray}
\label{eq: constrained TAP}
       &0=\frac{-m_i}{1-q}-\beta \partial_{m_i} H_J [\{m_j\}]-\frac{p_1(p_1-1)\beta^2 J_{p_1}^2}{2}(1-q)q^{p_1-2}m_i-\frac{p_2(p_2-1)\beta^2 J_{p_2}^2}{2}(1-q)q^{p_2-2}m_i \; .
\end{eqnarray}
To sum up, the constrained free energy describes a TAP state $\{m_i\}$ with a spherical norm $l=1$ and an overlap $\sum_i m_i v_i=N q_o$ with the reference. It is a metastable TAP state at the  temperature $1/\beta$ when the constrained free energy is extremised with respect to $q_o$.

\begin{figure}
    \centering
    \includegraphics[width=0.65\textwidth]{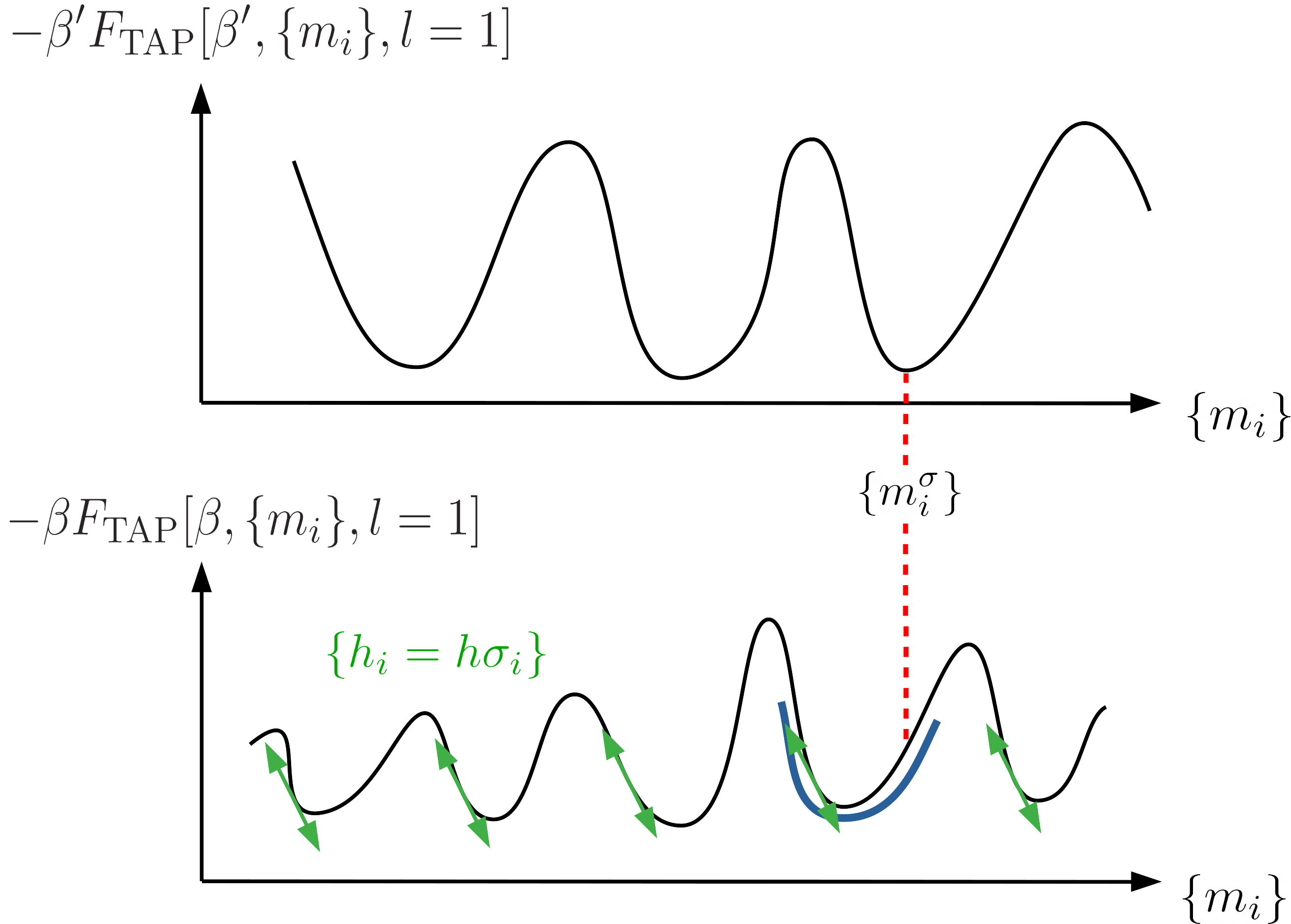}
    \includegraphics[width=0.61\textwidth]{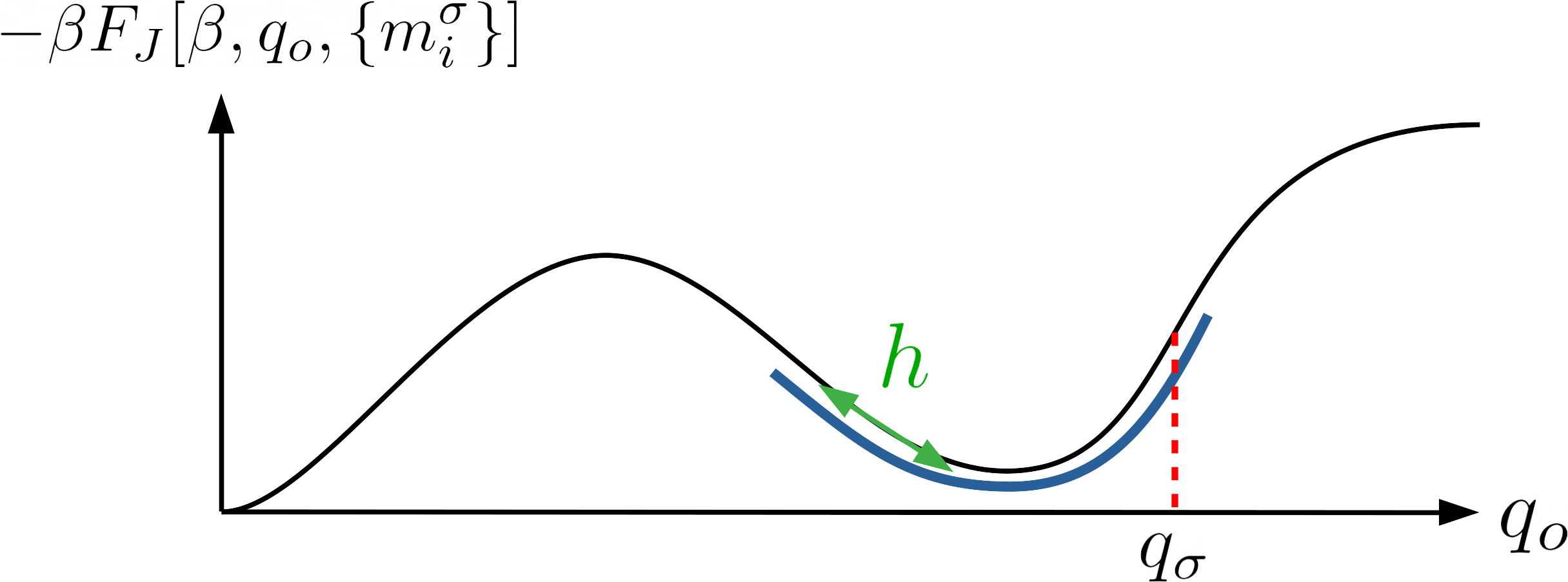}
    \caption{The upper and middle graphs represent the TAP free energy landscape at temperature $T'=1/\beta'$ and $T=1/\beta$. The lower graph represent the constrained free energy lanscape at temperature $T=1/\beta$. As the reference state plays a particular role in our discussion we emphasised its coordinates ($\{m_i^\sigma\}$ for the TAP free energy and $q_\sigma$ for the constrained one) with red dashed lines. If we select the convex/concave region around $q_o=q_\sigma$, the constrained free energy describes a TAP state in the same convex/concave region as the reference state $\{m_i^\sigma\}$. The two regions are delimited in blue in the middle and lower graphs. The fields $h$ and $\{h_i\}$ are defined through the slope of their respective free energy, we emphasised here the non-bijective nature of the Legendre transforms due to this definition with the green arrows. Indeed it is direct to see that a non convexe/concave function can give a same set of fields $\{h_i\}$ and $h$ for different values of $\{m_i\}$ and $q_o$. }
    \label{fig:quench}
\end{figure}

There is one last ambiguity that we have to take care of with the Legendre transforms. In fact we can map one value of $q_o$ with one value of $h$ only in a region where the constrained free energy is either convex or concave. The same problem appears with the TAP free energy for the conjugate variables $\{h_i\}$ and $\{m_i\}$. More practically if we set a value for $q_o$, and we consequently fix $h$, there are still numerous sets of magnetisations $\{m_i\}$ that verify Eq.~(\ref{eq:h vs magnetisations}).
However, if we focus on a reference $\{v_i=m_i^\sigma\}$ being a metastable TAP state at $\beta'$, it is possible to pin the right set of magnetisations for a given value of $q_o$. Indeed one can remember that the constrained free energy (\ref{eq:free energy mixed p-spin}) is known exactly under this assumption when $q_o=q_\sigma$:
\begin{equation}
          -\beta F_J[\beta,q_o=q_\sigma,\{m_i^{\sigma}\}] =-\beta F_{\rm TAP}\big[\beta,\{ m_i^{\sigma}\},l=1\big] \; .
\end{equation}
 In that case the constrained free energy describes a system with magnetisations $\{m_i^\sigma\}$ at  temperature $1/\beta$. Consequently, the constrained free energy in the convex/concave region around $q_o=q_\sigma$  describes a TAP state with magnetisation $\{m_i\}$ in the convex/concave region around $\{m_i^\sigma\}$ - see Fig.~\ref{fig:quench}. Like with a general reference state $\{v_i\}$, this TAP state has a spherical norm $l=1$ and an overlap $\sum_i m_i m_i^\sigma=N q_o$ with the reference. It also extremises the constrained free energy when it becomes metastable, in other words when the local magnetisations follow Eq.~(\ref{eq: constrained TAP}).
 
\section{Conclusions and outlook}
In this paper we first showed how chaos is present for any quench when the system is initially equilibrated at $T'\in [T_s,T_d]$. Using both the TAP approach and the FP potential we saw that there is in fact no simple rescaling $m_i \rightarrow\alpha \,m_i$ for all magnetisations linking the reference state at a given temperature to the constrained one at a different temperature. In dynamic terms, this is interpreted as a non trivial relation, or an absence of any simple one, between the initial state and the quenched one.
We then introduced a constrained free energy which enabled us to describe a system equilibrated at temperature $T$ enforced to have a fixed overlap with a reference state. Performing a high temperature expansion of this free energy we saw the role of the reference in the context of constrained equilibrium. In particular for the mixed $p$-spin model, each reference -taken metastable at a temperature $T'$- yields a different equilibrium state depending on the value taken by a ``parameter" that we called $\Delta H_{p_2}[\{m_i^\sigma\}]$. Finally we linked this new free energy to the unconstrained TAP one; it demonstrated that the equilibrated system corresponds to one given TAP state which is metastable when the constrained free energy is extremised with respect to the overlap with the reference.

To have here a complete understanding of the metastable properties the number of metastable TAP states (also called complexity) with fixed parameters $q$, $H_{p_1}[\{m_i\}]$, $H_{p_2}[\{m_i\}]$ and
$\Delta H_{p_2}[\{ m_i^\sigma \}]$ is still missing. This quantity would probably allow us to match the constrained free energy with the Franz-Parisi potential (up to $O(\beta^3)$ correction terms) in the same fashion that the TAP free energy is linked via an extra complexity term to the free energy derived with the replica trick. Besides this calculation would allow us to know in which interval is the value of $\Delta H_{p_2}[\{ m_i^\sigma \}]$ expected. 

Another interesting perspective would be to determine the complexity $\Sigma[q,q_o,\{m_i^\sigma\}]$ of the metastable TAP states with an order parameter $q$ and a fixed overlap $q_o$ with another metastable TAP state $\{m_i^\sigma\}$. Previous papers have already pushed forward in this direction. For example in Refs.~\cite{Ros2018,Ros2019} Ros {\it et al} have calculated such a complexity with an extra average over the disorder, in other words they performed the annealed average ${\rm I\!E}\Big[\exp{(\Sigma[q,q_o,\{m_i^\sigma\}]\big)}\Big]$. Unfortunately as our TAP-like approach is specifically disordered dependent we cannot use directly their results in our discussion.

\label{sec:co}

\acknowledgements{We warmly thank J. Kurchan, T. Rizzo,  M. Tarzia and F. Zamponi for very useful discussions and suggestions.}

\appendix

\section{TAP states}
\label{sec:app_TAP}

The metastable states of disordered models can be accessed with the Thouless-Anderson-Palmer (TAP) method \cite{Thouless1977}. We recall known results about these states here and we derive some new features for the mixed spherical model, of interest for our purposes. 

\subsection{Definitions and TAP free energy density}

The quenched randomness induces a complex free energy landscape with numerous saddle-points that are, with a generalisation of the Landau arguments, interpreted as metastable states. The free energy landscape is parametrised 
by a set of $N$ local order parameters, or local 
magnetisations, $\{m_i\}$ defined as 
\begin{equation}
    m_i \equiv \langle s_i \rangle 
    =\sqrt{q} \, \sigma_i \quad \forall i
\end{equation}
with 
\begin{equation}
    \sigma_i \in {\rm I\!R}\text{,}\qquad q \equiv \frac{1}{N}\sum_i {m_i}^2 \qquad\text{and}\qquad 
    \sum_i \sigma^2_i=lN
\end{equation}
in spherical models.
The local metastable states, or TAP states, are given by, on the one hand, the paramagnetic solution with all $m_i=0$ and, on the other hand, at sufficiently low temperatures, different sets of non-vanishing values of the $\{m_i\}$ that are extrema of the free energy density.

We define the adimensional energy density
\begin{equation}
  \varepsilon_{p_1}= -\frac{1}{N}\sum_{i_1  < \dots < i_{p_1}}  \frac{J_{i_1 \dots i_{p_1}} m_{i_1} \dots m_{i_{p_1}}}{J_{p_1} q^{\frac{p_1}{2}}}=-\frac{1}{N}\sum_{i_1  < \dots < i_{p_1}}  \frac{J_{i_1 \dots i_{p_1}}} {J_{p_1}}\sigma_{i_1} \dots \sigma_{i_{p_1}}
\end{equation}
and similarly for $\varepsilon_{p_2}$. These energy densities depend only on the local  magnetisation orientations $\{\sigma_i\}$ and not on the global order parameter $q$. In the metastable states we expect them to be negative.

Using the definitions above,  
the free energy density of a given TAP state can be parametrised in terms of the adimensional energetic contributions of the two terms in the Hamiltonian $(\varepsilon_{p_1},\varepsilon_{p_2})$ and the parameter $q$:
\begin{eqnarray}
    \label{eq:free_energy_TAP}
    F_{\rm TAP}[\beta,\{m_i\},l] &=&N f_{\rm TAP}[\beta,\{m_i\},l] =N f_{\rm TAP}(\beta,q,\varepsilon_{p_1},\varepsilon_{p_2},l)
\\
    &=&
    \; 
    -\frac{NT}{2}\ln(l-q)+ N q^{\frac{p_1}{2}} \, J_{p_1} \, \varepsilon_{p_1} +N q^{\frac{p_2}{2}} \, J_{p_2} \, \varepsilon_{p_2}
    \nonumber\\
    & &  -\frac{N J_{p_1}^2}{4T}\Big[l^{p_1}-p_1 l q^{p_1-1}+(p_1-1)q^{p_1}\Big]-\frac{N J_2^2}{4T}\Big[l^{p_2}-p_2 l q^{p_2-1}+(p_2-1)q^{p_2}\Big]  
    \; .    \nonumber
\end{eqnarray}
This form appeared in Eq.~(2) in~\cite{Annibale2004} and Eq.~(30) in~\cite{Rizzo2006}. 

\subsection{Reminder on the derivation of the TAP free energy density}
\label{sec: app TAP free energy}
We quickly recall here how the TAP free energy can be derived with a high temperature expansion. This approach, proposed in \cite{Biroli1999,Georges1991}, starts by considering the free energy
\begin{eqnarray}
\label{eq: free energy def2}
    -\beta F_{\rm TAP}[\beta,\{m_i\},l]&=&\ln \Bigg\{  {\rm Tr}_{\{s_i\}}\Big[e^{-\beta H_{J}[\{s_i\}]-\frac{\lambda}{2}\sum_i (s_i^2-l)-\sum_ih_i(s_i -m_i)}\Big]  \Bigg\}=\ln Z_{\rm TAP}[\beta,\{m_i\},l]\\
    &=&-\beta F_{\rm TAP}^\star[\beta,\{h_i\},\lambda]+\frac{N \lambda l}{2}+\sum_i h_i m_i \nonumber
 \end{eqnarray}
 up to order $O(N)$, with
 \begin{eqnarray}
   \partial_\lambda \big(-\beta F_{\rm TAP}[\beta,\{m_i\},l]\big)    &=& 0 \implies \sum_i \langle s_i^2 \rangle= N l \; ,
\\
       \partial_{h_i} \big(-\beta F_{\rm TAP}[\beta,\{m_i\},l]\big)    
       &=&    
       0 \implies \langle s_i \rangle =m_i \; ,
\end{eqnarray}
or
\begin{eqnarray}
       \partial_\lambda \big(-\beta F_{\rm TAP}^\star[\beta,\{h_i\},\lambda]\big)    
       &=&    
       \frac{N l}{2} \implies \sum_i \langle s_i^2 \rangle= N l \; ,
\\
       \partial_{h_i} \big(-\beta F_{\rm TAP}^\star[\beta,\{h_i\},\lambda]\big)    
       &=&    
       m_i \;\implies \sum_i \langle s_i \rangle =m_i \; .
 \end{eqnarray}
$F_{\rm TAP}^\star [\beta,\{h_i\},\lambda]$ is the Legendre transform of the free energy $F_{\rm TAP}[\beta,\{m_i\},l]$.

 A Taylor expansion of the free energy can be performed around $\beta=0$, the interest of this method is that extensive terms arise only up to $O(\beta^2)$. Therefore, in the large N limit, the series is truncated at such low order, allowing for an easy and systematic derivation of the free energy. Concretely speaking, the series reads
 \begin{eqnarray}
   -\beta F_{\rm TAP}[\beta,\{m_i\},l] & = & \sum_{k=0}^{+\infty}\frac{\beta^k}{k!}\partial^k_\beta (-\beta F_{\rm TAP})\Bigr|_{\beta=0}\\
   & = & -\beta F_{\rm TAP}\Bigr|_{\beta=0} +\beta \partial_\beta (-\beta F_{\rm TAP})\Bigr|_{\beta=0} +\frac{\beta^2}{2} \partial^2_\beta (-\beta F_{\rm TAP})\Bigr|_{\beta=0}+o(N)
   \; , \nonumber
 \end{eqnarray}
and the procedure yields in the end
 \begin{eqnarray}
    F_{\rm TAP}[\beta,\{m_i\},l] &=&
    \; 
    -\frac{NT}{2}\ln(l-q)+ N q^{\frac{p_1}{2}} \, J_{p_1} \, \varepsilon_{p_1} +N q^{\frac{p_2}{2}} \, J_{p_2} \, \varepsilon_{p_2}
    \\
    & &  -\frac{N J_{p_1}^2}{4T}\Big[l^{p_1}-p_1 l q^{p_1-1}+(p_1-1)q^{p_1}\Big]-\frac{N J_2^2}{4T}\Big[l^{p_2}-p_2 l q^{p_2-1}+(p_2-1)q^{p_2}\Big]  
    \;     \nonumber
\end{eqnarray}
with
 \begin{eqnarray}
       h_i&=&\frac{-m_i}{1-q}-\frac{\beta}{N} \partial_{m_i} H_J [\{m_j\}]-\frac{p_1(p_1-1)\beta^2 J_{p_1}^2}{2}(l-q)q^{p_1-2}m_i-\frac{p_2(p_2-1)\beta^2 J_{p_2}^2}{2}(l-q)q^{p_2-2}m_i \; ,\\
       \lambda&=&\frac{1}{1-q}+\frac{p_1\beta^2 J_{p_1}^2}{2}(l^{p_1-1}-q^{p_1-1})
\end{eqnarray}

In the following we will always consider the usual spherical constrain
\begin{eqnarray}
       \sum_i \langle s_i^2 \rangle =N \implies l=1.
\end{eqnarray}
\subsection{Paramagnetic solution}

The paramagnetic state is characterised by vanishing local magnetisations,  $m_i=0$, implying 
$q=0$ and 
\begin{equation}
    f_{\rm TAP}[\beta,\{m_i\},l=1]
    =
    -\frac{1}{4T} \, (J_{p_1}^2+ J_{p_2}^2)
    \; .
\end{equation}
Besides, this solution is stable as the Hessian 
$\frac{\partial^2 f_{\rm TAP}[\beta,\{m_i\},l=1]}{\partial m_i\partial m_j}\Bigr|_{m_i=0}$ 
has positive eigenvalues at all temperatures:
\begin{eqnarray}
  \frac{\partial^2 f_{\rm TAP}[\beta,\{m_i\},l=1]}{\partial m_i\partial m_j}\Bigr|_{m_i=0}=\delta_{ij}    \quad\text{for}\quad p_1,p_2\geq 3  \; .
\end{eqnarray}{}

\subsection{Elliptic solutions}
\label{sec:TAP_metastability}

Equation~(\ref{eq:free_energy_TAP}) is written as a function of the 
random variables $(\varepsilon_{p_1}, \varepsilon_{p_2})$ that themselves depend on the orientation of 
the $N$-dimensional vector 
${\bf \sigma}=(\sigma_1, \dots, \sigma_N)$ 
and the global parameter $q$. As all solutions with the same pair of contributions to the total energy $(\varepsilon_{p_1}, \varepsilon_{p_2})$ have the same $f_{\rm TAP}$,
it is convenient to start by studying this function at fixed $(\varepsilon_{p_1}, \varepsilon_{p_2})$ for varying $q$.

Let us therefore impose the extremal condition 
$\partial_q f_{\rm TAP}[\beta,\{m_i\},l=1] =0$, that implies 
\begin{equation}
    \begin{split}
        \frac{p_1(p_1-1)}{2}z_1^2+p_1 \varepsilon_{p_1} z_1+\frac{p_2(p_2-1)}{2}z_2^2+p_2 \varepsilon_{p_2} z_2+1=0
    \end{split}
    \label{eq:TAP2}
\end{equation}
with
\begin{equation}
    z_\ell \equiv \frac{J_{p_\ell}}{T} (1-q)q^{\frac{p_\ell}{2}-1}
    \qquad\qquad \mbox{for} \qquad \ell=1,2
    \; .
    \label{eq:zell-limit}
\end{equation}
This equation can be rewritten in a more convenient way using the definitions
\begin{equation}
    X_\ell \equiv z_\ell+\frac{\varepsilon_{p_\ell}}{p_\ell-1} 
        \qquad\qquad \mbox{for} \qquad \ell=1,2
        \; .
    \label{eq:Xidef}
\end{equation}
It then reads
\begin{equation}
    \label{eq:ellipseX}
    \frac{p_1(p_1-1)}{2}X_1^2+\frac{p_2(p_2-1)}{2}X_2^2=
    -1 + 
    \frac{p_1 \varepsilon_{p_1}^2 }{2 (p_1-1)}+\frac{p_2\varepsilon_{p_2}^2 }{2(p_2-1)} \equiv R^2 \; .
\end{equation}
One should keep in mind that $\varepsilon_{p_1}$ and $\varepsilon_{p_2}$ are not independent.

\subsubsection{Limit states}

A limiting relation between the two energies is derived by requiring that both the right-hand and left-hand side of Eq.~(\ref{eq:ellipseX}) vanish, that is to 
say, by imposing $R=0$. 
From the right-hand side of the equation one has
\begin{equation}
    1=\frac{p_1 \varepsilon_{p_1}^2 }{2(p_1-1)}+
    \frac{p_2 \varepsilon_{p_2}^2 }{2(p_2-1)}
    \; . 
    \label{eq:elliptic-energies}
\end{equation}
We will call the configurations that satisfy this equation {\it limit states}. They are represented by the dark limit of the quarter of ellipse in Fig.~\ref{fig:limit_state}(a). 
This is another elliptic equation, now seen as a function of $\varepsilon_{p_1}$ and $\varepsilon_{p_2}$. Energies within the shaded quarter in Fig.~\ref{fig:limit_state} are 
forbidden since they would lead to negative values of $R^2$. 

Coming back to the vanishing condition on the left-hand side of Eq.~(\ref{eq:ellipseX}), it implies $X_1=X_2=0$, and one then deduces
\begin{eqnarray}
    z_\ell=-\frac{\varepsilon_{p_\ell}}{p_\ell-1}
        \qquad\qquad \mbox{for} \qquad \ell=1,2
    \; .
    \label{limit_state_12}
\end{eqnarray}
Injecting now these expressions for $z_1$ and $z_2$ in Eq.~(\ref{eq:TAP2}),
we find
\begin{equation}
 \label{eq:q-threshold2}
 T^2 = \frac{p_1}{2} J_{p_1}^2 (p_1-1)q^{p_1-2} (1-q)^2 + \frac{p_2}{2} J_{p_2}^2 (p_2-1)q^{p_2-2} (1-q)^2 
\end{equation}
and, in a more compact notation,
\begin{equation}
 \label{eq:q-threshold}
 T^2 = \nu''(q) (1-q)^2  
 \; ,
\end{equation}
where the function $\nu$ is the Hamiltonian correlation defined in Eq.~(\ref{eq:random-pot-corr}) and the two primes indicate a double derivative with respect to the argument.
This equation is the same as the one obtained by requiring marginality in the replica analysis~\cite{Crisanti2006} and we will call it the marginality condition.
It determines $q$ at a given temperature for the states with energies $\varepsilon_{p_1}$ and $\varepsilon_{p_2}$ lying on the ellipse defined in Eq.~(\ref{eq:elliptic-energies}). The right-hand-side has the usual bell-shape form. The equation has solution
$q=1-a T$ with $a=T/\sqrt{\nu''(1)}$ for $T\to 0$ and it admits a physical solution, one with $q$ decreasing for increasing temperature, until a maximal temperature determined by the maximal value of the right-hand-side.

Let us now focus on the birth and disappearance temperature of these states. Using Eqs.~(\ref{eq:zell-limit}) and~(\ref{limit_state_12}) one  straightforwardly obtains
\begin{equation}
    {\varepsilon_{p_1}^{ell}} =\frac{J_{p_1}(p_1-1)}{J_{p_2}(p_2-1)} \, q^{\frac{p_1-p_2}{2}} \, {\varepsilon_{p_2}^{ell}} 
    \; .
    \label{eq:linear-e1e2}
\end{equation}
This is a straight line going through the origin, with a positive slope controlled by $q$ and, therefore, by $T$ through Eq.~(\ref{eq:q-threshold}). Since $q$ takes, at most, the value $q=1$ at $T=0$, the maximum slope is $J_{p_1} (p_1-1)/[J_{p_2} (p_2-1)]$. On the other hand, $q$ cannot be smaller than $q_{\rm max}$ (the value of $q$ at which the bell-shaped curve reaches its maximum). Therefore, the minimal slope is $J_{p_1}(p_1-1) /[J_{p_2}(p_2-1)] \; q_{\rm max}^{{p_1-p_2}/{2}}$. This argument sets the two limiting green straight lines in Fig.~\ref{fig:limit_state}(a) and proves that the only allowed states on the ellipse have energies on the dashed (green) arc.

We want to understand next which are the energies of the limit states. Working with the ellipse equation~(\ref{eq:elliptic-energies}) and the linear relation between the angular energies modulated by a factor that depends on $q$, Eq.~(\ref{eq:linear-e1e2}), one derives
\begin{equation}
    \varepsilon_{p_\ell} = - [\nu''(q)]^{-1/2} J_{p_\ell} (p_\ell-1) q^{(p_\ell-2)/2}
    \; . 
    \label{eq:e1e2limit}
\end{equation}

The total energy density of a limit state, 
given by the first four terms in $f_{\rm TAP}$ evaluated at $\varepsilon_{p_1}$ and $\varepsilon_{p_2}$ in Eq.~(\ref{eq:e1e2limit}) is then
\begin{eqnarray}
    e_{\rm TAP} &=& 
     - \left[ \nu''(q) \right]^{-1/2} \sum_\ell J^2_{p_\ell} (p_\ell-1) q^{p_\ell-1}
     - \frac{1}{2T} \sum_{\ell} J_{p_\ell}^2
     \left[ 1- p_\ell q^{p_\ell-1} + (p_\ell-1) q^{p_\ell}
     \right]
     \nonumber\\
     &=&-\frac{1}{2T} 
     \sum_\ell J_{p_\ell}^2
    [1+(p_\ell-2) q^{p_\ell-1} -
     (p_\ell-1) q^{p_\ell}]\; .
\end{eqnarray}

\subsubsection{Marginal states, with $q$ given by the marginality equation}

\begin{figure}[h]
    \centering
    \includegraphics[width=0.47\textwidth]{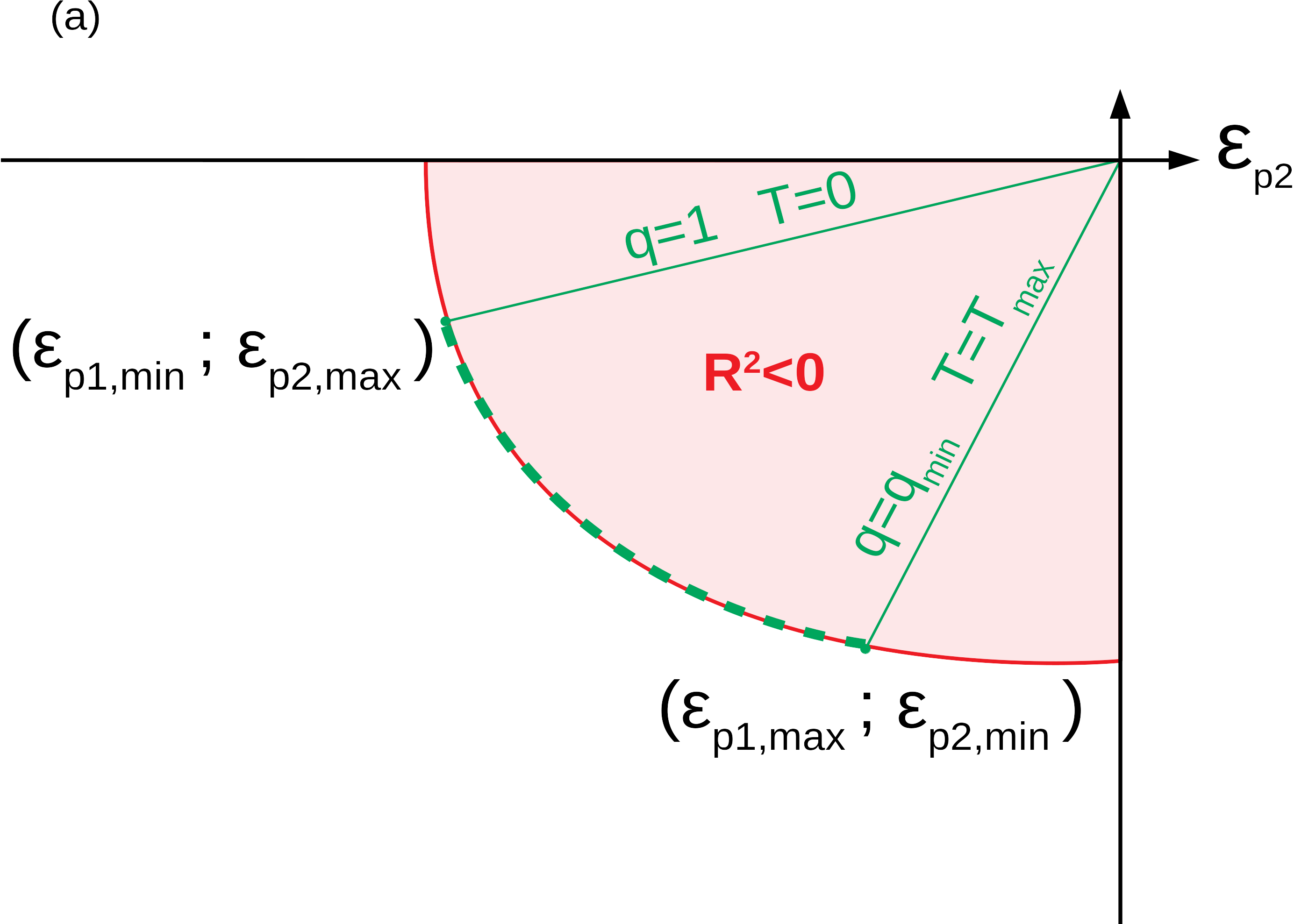}
    \includegraphics[width=0.47\textwidth]{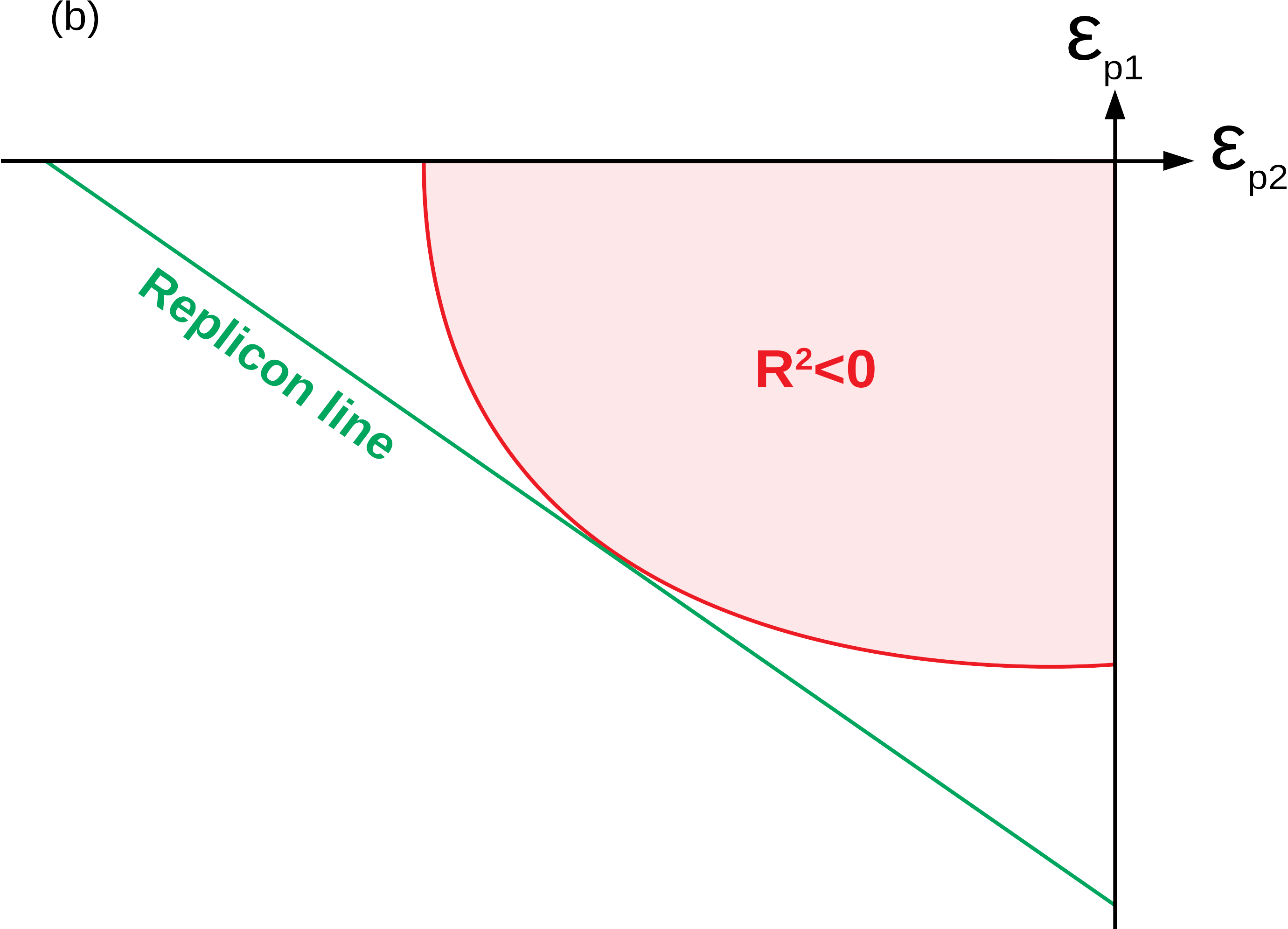}
    \caption{The red thick arc corresponds to the adimensional energy densities $\varepsilon_{p_1}$ and $\varepsilon_{p_2}$ satisfying $R^2=0$, see Eq.~(\ref{eq:elliptic-energies}). Inside this limit curve (shaded red zone) there are no TAP states. In panel (a) the dashed green line on the arc signals where are located all the limit states accessible from $T=0$ to $T=T_{max}$. In panel (b) the states following the replicon condition~(\ref{eq:q-threshold}) are all on a straight line (green here) tangent to the limiting curve.
    }
    \label{fig:limit_state}
\end{figure}

If we now fix the temperature and consider that $q$ is determined by Eq.~(\ref{eq:q-threshold}), the $z_1$ and $z_2$ values are also fixed, and Eq.~(\ref{eq:TAP2}) yields a linear relation between the two 
angular energies $\varepsilon_{p_1}$ and $\varepsilon_{p_2}$, with negative slope as well as negative intersection with the vertical axis:
\begin{equation}
\begin{split}
\label{linear_energies}
    p_1 \varepsilon_{p_1} & = 
    - \frac{z_2}{z_1}p_2 \varepsilon_{p_2} -\frac{1}{z_1}
    -\sum_{\ell=1}^2 \frac{p_\ell (p_\ell-1) z_\ell^2}{2z_1}
    =- \frac{z_2}{z_1}p_2 \varepsilon_{p_2} -\frac{2}{z_1}\; ,
\end{split}
\end{equation}
that is shown with a (green) straight line and called replicon line in Fig.~\ref{fig:limit_state}(b).
A geometric argument shows that this (green) straight line must be tangent to the (red) ellipse at the point $(\varepsilon_{p_1},\varepsilon_{p_2})$ with coordinates
\begin{equation}
\begin{split}
  & \varepsilon_{p_\ell}^{ell}=-  (p_\ell-1) z_\ell
  \; .
\end{split}
\end{equation}
In fact,
for a given $T$, Eq.~(\ref{eq:q-threshold2}) fixes $q$, and Eq.~(\ref{linear_energies}) determines all the energy densities ($\varepsilon_{p_1},\varepsilon_{p_2}$) that are in principle possible 
for this pair ($T,q$). The set of ($\varepsilon_{p_1},\varepsilon_{p_2}$) thus determined should include 
the densities (${\varepsilon_{p_1}^{ell}},{\varepsilon_{p_2}^{ell}}$) 
that lie on 
the ellipse. Geometrically, the only way to approach this point with a straight line without crossing the limit curve and getting inside the forbidden shaded red zone is to take its tangent.
Therefore, the energy densities corresponding to pairs ($T,q$) linked by Eq.~(\ref{eq:q-threshold2}) lie on a green straight line as the one drawn in Fig.~\ref{fig:limit_state}(b). At a different temperature the slope of the straight line 
and touching point on the ellipse will be different.

Another way to see that the straight (green) line should be tangent to the limit curve is to calculate the infinitesimal variation of $R^2$ around the 
point (${\varepsilon_{p_1}^{ell}},{\varepsilon_{p_2}^{ell}}$). Taking the ellipse equation~(\ref{eq:ellipseX}) this variation is given by
\begin{equation}
\begin{split}
   d\Big[R^2\Big]&=p_1 (p_1-1)X_1 dX_1+p_2 (p_2-1)X_2 dX_2.\\
\end{split}
\end{equation}
Considering a point on the limit curve gives $X_1=X_2=0$, thus we get $d\Big[R^2\Big]=0$.
This result implies that any first order variation around a point from the limit curve shall keep $R^2$ constant, the only straight line verifying this property is the tangent to the curve.

\section{Details on the Taylor expansion of the free energy}
\label{sec:appA}

In this Appendix we will detail some calculation steps leading to the free energy in Eq.~(\ref{eq:free energy}). For more clarity, we will consider each term in Eq.~(\ref{eq:2nd order free energy term}) 
separately.

To determine the extensive contribution to the free energy of the variance $\Big\langle {H_J[\{s_i\}]}^2 \Big\rangle-\Big\langle H_J[\{s_i\}]\Big\rangle^2$ we follow arguments given by Rieger in Ref.~\cite{Rieger1992}. We start by  separating different
contributions to the square:

\begin{eqnarray}
\label{eq: AppB1}
      \Big(\sum_{\{i_{p_1}\}}J_{\{i_{p_1}\}} S_{\{i_{p_1}\}}\Big)^2 & = &\sum_{ \{i_{p_1}\} \neq\{j_{p_1}\}} {J_{\{i_{p_1}\}}}{J_{\{j_{p_1}\}}} {S_{\{i_{p_1}\}}}{S_{\{j_{p_1}\}}}\nonumber\\
      &&+\sum_{\{i_{p_1}\}} J_{\{i_{p_1}\}}^2 {S_{\{i_{p_1}\}}}^2 \nonumber\\
      & & +\sum_k \sum_{\substack{ \{i_2,i_{p_1}\} \\ \neq\\ \{j_2,j_{p_1}\}}} J_{k,\{i_{2},i_{p_1}\}} J_{k,\{j_{2},j_{p_1}\}} {s_{k}}^2 S_{\{i_{2},i_{p_1}\}}S_{\{j_{2},j_{p_1}\}} \nonumber\\
      & & +\sum_{k,l} \sum_{\substack{ \{i_3,i_{p_1}\} \\ \neq\\ \{j_3,j_{p_1}\}}} J_{k,l,\{i_{3},i_{p_1}\}} J_{k,l,\{j_{3},j_{p_1}\}} {s_{k}}^2 {s_{l}}^2 S_{\{i_{3},i_{p_1}\}}S_{\{j_{3},j_{p_1}\}} +o(N).
\end{eqnarray}
We considered here that the terms left out are sub-extensive.
This assumption will be proven in the following discussion. The first sum is irrelevant for our calculation as it appears both in $\Big\langle {H_J[\{s_i\}]}^2 \Big\rangle$ and $\Big\langle {H_J[\{s_i\}]} \Big\rangle^2$; thus, 
it cancels after performing the average. The second term in Eq.~(\ref{eq: AppB1}) 
is extensive, in fact $J_{\{i_{p_1}\}}^2 \sim O(N^{-p_1+1})$ and thus $\sum_{\{i_{p_1}\}} O(N^{-p_1+1}) \sim N^{p1} O(N^{-p_1+1}) \sim O(N)$. The third one yields also an extensive contribution; looking more carefully at it we have $J_{k,\{i_{2},i_{p_1}\}} J_{k,\{j_{2},j_{p_1}\}} \sim O(N^{-p_1+1}) \times O(N^{-p_1+1})$ and then $\sum_k \sum_{\substack{ \{i_2,i_{p_1}\} \\ \neq\\ \{j_2,j_{p_1}\}}} J_{k,\{i_{2},i_{p_1}\}} J_{k,\{j_{2},j_{p_1}\}} \sim  N\times  N^{2 p_1-2} O(N^{-2 p_1+2}) \sim O(N)$. The last term is the first one which does not give an extensive contribution as $\sum_{k,l} \sum_{\substack{ \{i_3,i_{p_1}\} \\ \neq\\ \{j_3,j_{p_1}\}}} J_{k,l,\{i_{3},i_{p_1}\}} J_{k,l,\{j_{3},j_{p_1}\}} \sim  N^2 \times  N^{2 p_1-4} O(N^{-2 p_1+2}) \sim O(N^{-1})$. 
Following the estimation of this last sum it is then straightforward to see why the left out terms are not extensive. 

In the same fashion it can also be shown that the extensive contributions between the crossed terms $J_{p_1}-J_{p_2}$ are
\begin{eqnarray}
       \sum_{\{i_{p_1}\},\{j_{p_2}\}}J_{\{i_{p_1}\}}J_{\{j_{p_2}\}} S_{\{i_{p_1}\}}S_{\{i_{p_2}\}} = \sum_k \sum_{\substack{ \{i_2,i_{p_1}\} \\ \{j_2,j_{p_2}\}}} J_{k,\{i_{2},i_{p_1}\}} J_{k,\{j_{2},j_{p_2}\}} {s_{k}}^2 S_{\{i_{2},i_{p_1}\}}S_{\{j_{2},j_{p_2}\}} 
       \; . 
\end{eqnarray}
From this discussion it is now possible to properly calculate $\Big\langle {H_J[\{s_i\}]}^2 \Big\rangle-\Big\langle H_J[\{s_i\}]\Big\rangle^2$.
In fact,
\begin{eqnarray}
    \Big\langle {H_J[\{s_i\}]}^2 \Big\rangle-\Big\langle H_J[\{s_i\}]\Big\rangle^2 & = & \sum_{\{i_{p_1}\}} J_{\{i_{p_1}\}}^2 \Delta S_{\{i_{p_1}\}}+\sum_{\{i_{p_2}\}} J_{\{i_{p_2}\}}^2 \Delta S_{\{i_{p_2}\}} 
    \nonumber\\
    & &+\sum_k \sum_{\substack{ \{i_2,i_{p_1}\} \\ \neq\\ \{j_2,j_{p_1}\}}} J_{k,\{i_{2},i_{p_1}\}} J_{k,\{j_{2},j_{p_1}\}} \Big( \langle {s_{k}}^2 \rangle - \langle s_{k} \rangle^2\Big) \langle S_{\{i_{2},i_{p_1}\}}\rangle \langle S_{\{j_{2},j_{p_1}\}} \rangle
    \nonumber\\
    & &+2\sum_k \sum_{\substack{ \{i_2,i_{p_1}\} \\ \{j_2,j_{p_1}\}}} J_{k,\{i_{2},i_{p_1}\}} J_{k,\{j_{2},j_{p_2}\}} \Big( \langle {s_{k}}^2 \rangle - \langle s_{k} \rangle^2\Big) \langle S_{\{i_{2},i_{p_1}\}}\rangle \langle S_{\{j_{2},j_{p_2}\}} \rangle
    \nonumber\\
    & &+\sum_k \sum_{\substack{ \{i_2,i_{p_2}\} \\ \neq\\ \{j_2,j_{p_2}\}}} J_{k,\{i_{2},i_{p_2}\}} J_{k,\{j_{2},j_{p_2}\}} \Big( \langle {s_{k}}^2 \rangle - \langle s_{k} \rangle^2\Big) \langle S_{\{i_{2},i_{p_2}\}}\rangle \langle S_{\{j_{2},j_{p_2}\}} \rangle 
    \; .
\end{eqnarray}
This expression can be transformed into
\begin{eqnarray}
    \Big\langle {H_J[\{s_i\}]}^2 \Big\rangle-\Big\langle H_J[\{s_i\}]\Big\rangle^2 & = & \sum_{\{i_{p_1}\}} J_{\{i_{p_1}\}}^2 \Delta S_{\{i_{p_1}\}}+\sum_{\{i_{p_2}\}} J_{\{i_{p_2}\}}^2 \Delta S_{\{i_{p_2}\}} 
    \nonumber\\
    & &-\Big(1-\frac{q_o^2}{q_\sigma}\Big) \sum_k \sum_{\{i_2,i_{p_1}\}} {J_{k,\{i_{2},i_{p_1}\}}}^2  \langle S_{\{i_{2},i_{p_1}\}}\rangle^2 
    \nonumber\\
    & &-\Big(1-\frac{q_o^2}{q_\sigma}\Big) \sum_k \sum_{\{i_2,i_{p_2}\}} {J_{k,\{i_{2},i_{p_2}\}}}^2  \langle S_{\{i_{2},i_{p_2}\}}\rangle^2 
    \nonumber\\
    & &+\Big(1-\frac{q_o^2}{q_\sigma}\Big)\sum_k \Big[\sum_{ \{i_2,i_{p_1}\}} J_{k,\{i_{2},i_{p_1}\}} \langle S_{\{i_{2},i_{p_1}\}}\rangle
    \nonumber\\
    & & \hspace{2.4cm} + \sum_{ \{j_2,j_{p_2}\}} J_{k,\{j_{2},j_{p_2}\}} \langle S_{\{j_{2},j_{p_2}\}}\rangle \Big]^2. 
\end{eqnarray}
This leads straightforwardly to Eq.~(\ref{eq: 2nd order TAP(1)}) which is later used to determine the constrained free energy.
Focusing now on the two remaining terms in Eq.~(\ref{eq:2nd order free energy term}), we derive 
 \begin{eqnarray}
\Bigg\langle\Big\{\sum_i(s_i m_i^\sigma-q_o)\Big\}^2\Bigg\rangle & = &
\Big\langle\sum_i(s_i m_i^\sigma-q_o)\sum_j(s_j m_j^\sigma-q_o)\Big\rangle\nonumber\\
& = &
\Big\langle\sum_i(s_i m_i^\sigma-q_o)^2\Big\rangle+\sum_{i\neq j}\langle s_i m_i^\sigma-q_o\rangle\langle s_j m_j^\sigma-q_o\rangle\nonumber\\
& = & 
\sum_i \Big( \langle {s_i}^2 \rangle - \langle s_i \rangle^2\Big){m_i^{\sigma}}^2 
\;  \underset{\beta = 0}{=} \;  N 
\Big(1-\frac{q_o^2}{q_\sigma}\Big)q_\sigma
\end{eqnarray}
and
 \begin{eqnarray}
      \Big\langle H_J[\{s_i\}]\sum_i(s_i m_i^\sigma-q_o)\Big\rangle
      & = &\nonumber
      p_1 \sum_{\{i_{p_1}\}}J_{\{i_{p_1}\}} \Big\langle s_{i_1}(s_{i_1} m_{i_1}^\sigma-q_o)\Big\rangle \langle S_{\{i_{2},i_{p_1}\}}\rangle\\
      & & 
      + \sum_{\{i_{p_1}\}}J_{\{i_{p_1}\}} \langle S_{\{i_{1},i_{p_1}\}}\rangle  \sum_{j\neq \{i_{p_1}\}}\langle s_j m_j^\sigma-q_o\rangle  
      \nonumber\\
      &  &
      + p_2 \sum_{\{i_{p_2}\}}J_{\{i_{p_2}\}} \Big\langle s_{i_1}(s_{i_1} m_{i_1}^\sigma-q_o)\Big\rangle \langle S_{\{i_{2},i_{p_2}\}}\rangle\nonumber\\
      & & 
      +\sum_{\{i_{p_2}\}}J_{\{i_{p_2}\}} \langle S_{\{i_{1},i_{p_2}\}}\rangle  \sum_{j\neq \{i_{p_2}\}}\langle s_j m_j^\sigma-q_o\rangle  \nonumber\\
      & = & 
      p_1 \sum_{\{i_{p_1}\}}J_{\{i_{p_1}\}} \Big( \langle {s_{i_1}}^2 \rangle - \langle s_{i_1} \rangle^2\Big){m_{i_1}^{\sigma}} \langle S_{\{i_{2},i_{p_1}\}}\rangle \nonumber\\
      & &+ p_2 \sum_{\{i_{p_2}\}}J_{\{i_{p_2}\}} \Big( \langle {s_{i_1}}^2 \rangle - \langle s_{i_1} \rangle^2\Big){m_{i_1}^{\sigma}} \langle S_{\{i_{2},i_{p_2}\}}\rangle \nonumber\\
      & \underset{\beta = 0}{=} & 
      p_1 \Big(1-\frac{q_o^2}{q_\sigma}\Big) \Big(\frac{q_o}{q_\sigma}\Big)^{p_1-1}H_{p_1}[\{ m_i^\sigma \}]   
      + p_2 \Big(1-\frac{q_o^2}{q_\sigma}\Big) \Big(\frac{q_o}{q_\sigma}\Big)^{p_2-1}H_{p_2}[\{ m_i^\sigma \}] \; .
 \end{eqnarray}
We used these two terms in Eqs.~(\ref{eq: 2nd order TAP(2)}) and (\ref{eq: 2nd order TAP(3)}).

 \section{Simplifying the constrained free energy when the reference is a metastable TAP state.}
 \label{sec:appB}
 
The metastable TAP states at temperature $T'=1/\beta'$ verify
 \begin{eqnarray}
 \label{eq: TAP eq}
    \frac{m_k^\sigma}{1-q_\sigma}&=&\beta'\sum_{\{i_2,i_{p_1}\}} J_{k,\{i_{2},i_{p_1}\}} M_{\{i_{2},i_{p_1}\}}+\beta'\sum_{\{i_2,i_{p_2}\}} J_{k,\{i_{2},i_{p_2}\}} M_{\{i_{2},i_{p_2}\}}
    \nonumber\\
    &&-{\beta'}^2 J_{p_1}^2(1-q_\sigma)\frac{p_1(p_1-1)}{2} {q_\sigma}^{p_1-2} m_k^\sigma -{\beta'}^2J_{p_2}^2(1-q_\sigma)\frac{p_2(p_2-1)}{2} {q_\sigma}^{p_2-2} m_k^\sigma
\end{eqnarray}
and
\begin{eqnarray}
 \label{eq: TAP eq(2)}
    \frac{Nq_\sigma}{1-q_\sigma}&=&-\beta' p_1 H_{p_1} [\{ m_i^\sigma \}]-\beta'p_2 H_{p_2} [\{ m_i^\sigma \}]
    \nonumber\\
    &&-{\beta'}^2 N (1-q_\sigma)\frac{p_1(p_1-1)}{2} {q_\sigma}^{p_1-1} -{\beta'}^2 N (1-q_\sigma)\frac{p_2(p_2-1)}{2} {q_\sigma}^{p_2-1} 
    \; .
\end{eqnarray}
Thus, one can rewrite the two last terms of the constrained free energy (\ref{eq:free energy(1)}) in the following way:
\begin{eqnarray}
      \sum_k \Big[\sum_{ \{i_2,i_{p_1}\}} J_{k,\{i_{2},i_{p_1}\}} \Big(\frac{{q_o}}{q_\sigma}\Big)^{p_1-1} M_{\{i_{2},i_{p_1}\}}+ \sum_{ \{j_2,j_{p_2}\}} J_{k,\{j_{2},j_{p_2}\}} \Big(\frac{{q_o}}{q_\sigma}\Big)^{p_2-1} M_{\{j_{2},j_{p_2}\}} \Big]^2 \hspace{2cm}
      \nonumber\quad\quad\quad \\
      =\sum_k \Bigg\{m_k^\sigma\Big[\frac{1}{\beta'(1-q_\sigma)}+{\beta'}(1-q_\sigma)\frac{p_1(p_1-1)}{2} {q_\sigma}^{p_1-2} +{\beta'}(1-q_\sigma)\frac{p_2(p_2-1)}{2} {q_\sigma}^{p_2-2} \Big]\Big(\frac{q_o}{q_\sigma}\Big)^{p_1-1}
      \nonumber\\
      +\Big[\Big(\frac{{q_o}}{q_\sigma}\Big)^{p_2-1}-\Big(\frac{{q_o}}{q_\sigma}\Big)^{p_1-1}\Big] \sum_{ \{j_2,j_{p_2}\}} J_{k,\{j_{2},j_{p_2}\}} M_{\{j_{2},j_{p_2}\}} \Bigg\}^2 \; , 
\end{eqnarray}
\begin{eqnarray}
       \frac{1}{ q_\sigma N} \Bigg\{ p_1 \Big(\frac{q_o}{q_\sigma}\Big)^{p_1-1} H_{p_1} [\{ m_i^\sigma \}] + p_2 \Big(\frac{q_o}{q_\sigma}\Big)^{p_2-1} H_{p_2} [\{ m_i^\sigma \}]\Bigg\}^2\hspace{6.4cm}
       \nonumber\\
      = \frac{1}{ q_\sigma N}\Bigg\{ N\Big[\frac{-q_\sigma}{\beta'(1-q_\sigma)}-{\beta'}(1-q_\sigma)\frac{p_1(p_1-1)}{2} {q_\sigma}^{p_1-1} -{\beta'}(1-q_\sigma)\frac{p_2(p_2-1)}{2} {q_\sigma}^{p_2-1} \Big]\Big(\frac{q_o}{q_\sigma}\Big)^{p_1-1}\nonumber\\
      +\Big[\Big(\frac{{q_o}}{q_\sigma}\Big)^{p_2-1}-\Big(\frac{{q_o}}{q_\sigma}\Big)^{p_1-1}\Big] p_2H_{p_2}[\{ m_i^\sigma \}]\Bigg\}^2
      \; .
\end{eqnarray}
The difference of these two terms -that will be called A- yields
\begin{eqnarray}
       A&=&\Bigg[\Big(\frac{{q_o}}{q_\sigma}\Big)^{p_2-1}-\Big(\frac{{q_o}}{q_\sigma}\Big)^{p_1-1}\Bigg]^2\Bigg[\sum_k \Big(\sum_{ \{j_2,j_{p_2}\}} J_{k,\{j_{2},j_{p_2}\}} M_{\{j_{2},j_{p_2}\}} \Big)^2-\frac{{p_2}^2}{N}{H_2 [\{ m_i^\sigma \}]}^2\Bigg]
       \nonumber\\
       &=&\Bigg[\Big(\frac{{q_o}}{q_\sigma}\Big)^{p_2-1}-\Big(\frac{{q_o}}{q_\sigma}\Big)^{p_1-1}\Bigg]^2\Bigg[\sum_k \Big(\partial_{m_k^\sigma} H_{p_2} [\{ m_i^\sigma \}] \Big)^2-\frac{{p_2}^2}{N}{H_2 [\{ m_i^\sigma \}]}^2\Bigg]\nonumber\\
       &=&\Bigg[\Big(\frac{{q_o}}{q_\sigma}\Big)^{p_2-1}-\Big(\frac{{q_o}}{q_\sigma}\Big)^{p_1-1}\Bigg]^2\Delta H_{p_2}[\{ m_i^\sigma \}]
       \; .
\end{eqnarray}
Moreover in order to study the stability of the system -with $\partial_{q_o}(-\beta F)$- one can focus on $\partial_{q_o} \Big[\big(1-\frac{q_o^2}{q_\sigma}\big) A\Big]$:
\begin{eqnarray}
       && \partial_{q_o} \Big[\big(1-\frac{q_o^2}{q_\sigma}\big) A\Big]
       = 
       \frac{-2 q_o}{q_\sigma}A+2\big(1-\frac{q_o^2}{q_\sigma}\big)\Delta H_{p_2}[\{ m_i^\sigma \}]
       \Bigg[\Big(\frac{{q_o}}{q_\sigma}\Big)^{p_2-1}-\Big(\frac{{q_o}}{q_\sigma}\Big)^{p_1-1}\Bigg]\Bigg[(p_2-1)\Big(\frac{{q_o}^{p_2-2}}{{q_\sigma}^{p_2-1}}\Big)-(p_1-1)\Big(\frac{{q_o}^{p_1-2}}{{q_\sigma}^{p_1-1}}\Big)\Bigg]
       \nonumber\\
       & & 
       \qquad{=} \; 
       \frac{-2 q_o}{q_\sigma}A+2\frac{\Delta H_{p_2}[\{ m_i^\sigma \}]}{\beta^2\big(1-\frac{q_o^2}{q_\sigma}\big)}  
       \Bigg[\frac{q_o z_2}{J_{p_2}{q_\sigma}^{\frac{p_2}{2}}}-\frac{q_o z_1}{J_{p_1}{q_\sigma}^{\frac{p_1}{2}}}\Bigg]\Bigg[(p_2-1)\frac{ z_2}{J_{p_2}{q_\sigma}^{\frac{p_2}{2}}}-(p_1-1)\frac{ z_1}{J_{p_1}{q_\sigma}^{\frac{p_1}{2}}}\Big)\Bigg]
       \nonumber\\
       & & 
       \qquad = \; 
       2 \frac{\Delta H_{p_2}[\{ m_i^\sigma \}]}{\beta^2\big(1-\frac{q_o^2}{q_\sigma}\big)}   \Bigg[\displaystyle{\frac{q_o z_2}{J_{p_2}{q_\sigma}^{\frac{p_2}{2}}}-\frac{q_o z_1}{J_{p_1}{q_\sigma}^{\frac{p_1}{2}}}}\Bigg]  \Bigg[\frac{ z_2}{J_{p_2}{q_\sigma}^{\frac{p_2}{2}}}\Big(p_2-1-\frac{1}{\frac{q_\sigma}{q_o^2}-1}\Big)-\frac{z_1}{J_{p_1}{q_\sigma}^{\frac{p_1}{2}}}\Big(p_1-1-\frac{1}{\frac{q_\sigma}{q_o^2}-1}\Big)\Bigg]
\end{eqnarray}
with
\begin{eqnarray}
       z_\ell=\frac{J_{p_\ell}}{T}\Big(1-\frac{q_o^2}{q_\sigma}\Big) \Big(\frac{q_o^2}{q_\sigma}\Big)^{\frac{p_\ell}{2}-1} \qquad\qquad \mbox{for} \qquad \ell=1,2
       \; . 
\end{eqnarray}

 \section{Link between the constrained and usual TAP free energies.}
 \label{sec:app_link_TAP_constrained}
We now detail how the constrained and TAP free energies can be exactly equal to each other through the high temperature development. We will focus on the mixed $p$-spin spherical model using two assumptions, the reference $\{v_i=m_i^\sigma\}$ has to be a metastable TAP state at $\beta'$ and the overlap $q_o$ set to $q_o=q_\sigma$. In the case of the pure model the second assumption ($q_o=q_\sigma$) is not necessary. As derived in Sec.~\ref{sec:Application to the mixed $p$-spin model} the constrained free energy is
\begin{eqnarray}
\label{eq: appD1}
        -\beta F_{ J}\big[\beta,q_o,\{m_i^{\sigma}\}\big] =-\beta F_{\rm TAP}\Big[\beta,\Big\{\frac{q_o}{q_\sigma} m_i^{\sigma}\Big\}\Big]+\frac{\beta^2}{2}\Big(1-\frac{q_o^2}{q_\sigma}\Big)\Bigg[\Big(\frac{{q_o}}{q_\sigma}\Big)^{p_2-1}-\Big(\frac{{q_o}}{q_\sigma}\Big)^{p_1-1}\Bigg]^2\Delta H_{p_2}\big[\{ m_i^\sigma \}\big]+O(\beta^3)\; .
\end{eqnarray}
One shall note that the reference has to be a metastable TAP state to derive this formula. Using now the assumption $q_o=q_\sigma$ it straightforwardly simplifies to
\begin{eqnarray}
\label{eq: appD2}
        -\beta F_{ J}\big[\beta,q_o=q_\sigma,\{m_i^{\sigma}\}\big] =-\beta F_{\rm TAP}\big[\beta,\{ m_i^{\sigma}\}\big]+O(\beta^3)
\end{eqnarray}
and
 \begin{eqnarray}
       h&=&\frac{1}{N}\partial_{q_o}\Big(-\beta F_{ J}\big[\beta,q_o,\{m_i^{\sigma}\}\big]\Big)\Bigr|_{q_o=q_\sigma}
       \\
       &=&\frac{-1}{1-q_\sigma}-\frac{\beta}{N q_\sigma} p_1 H_{p_1} \big[\{m_j^{\sigma}\}\big]-\frac{\beta}{N q_\sigma} p_2 H_{p_2} \big[\{m_j^{\sigma}\}\big]
       \nonumber\\
       &&-\frac{p_1(p_1-1)\beta^2 J_{p_1}^2}{2}(1-q_\sigma)q_\sigma^{p_1-2}-\frac{p_2(p_2-1)\beta^2 J_{p_2}^2}{2}(1-q_\sigma)q_\sigma^{p_2-2} +O(\beta^3)\; ,\nonumber
       \\
       \lambda&=&\frac{2}{N}\partial_{l}\Big(-\beta F_{ J}\big[\beta,q_o,\{m_i^{\sigma}\}\big]\Big)\Bigr|_{q_o=q_\sigma}
       \\
       &=&\frac{1}{1-q_\sigma}+\frac{p_1\beta^2 J_{p_1}^2}{2}(1-q_\sigma^{p_1-1})+\frac{p_2\beta^2 J_{p_2}^2}{2}(1-q_\sigma^{p_2-1}) +O(\beta^3)\; .\nonumber
 \end{eqnarray}
The first step is to reformulate the Lagrange multiplier term $h\sum_i (s_i m_i^\sigma-q_\sigma)$ by taking into account the expression of the variable $h$. In fact it can be rewritten
\begin{eqnarray}
\label{eq: app_constrains}
        h\sum_i (s_i m_i^\sigma-q_\sigma) &=&\sum_i (s_i m_i^\sigma-q_\sigma)\Big[\frac{-1}{1-q_\sigma}-\frac{\beta}{N q_\sigma} p_1 H_{p_1} \big[\{m_j^{\sigma}\}\big]-\frac{\beta}{N q_\sigma} p_2 H_{p_2} \big[\{m_j^{\sigma}\}\big]
       \\
       &&-\frac{p_1(p_1-1)\beta^2 J_{p_1}^2}{2}(1-q_\sigma)q_\sigma^{p_1-2}-\frac{p_2(p_2-1)\beta^2 J_{p_2}^2}{2}(1-q_\sigma)q_\sigma^{p_2-2} +O(\beta^3)\Big]\; 
       \nonumber\\
       &=&\sum_i s_i m_i^\sigma \Big[ -\frac{ 1}{1-q_\sigma} -\frac{\beta}{N q_\sigma} p_1 H_{p_1} \big[\{m_j^{\sigma}\}\big]-\frac{\beta}{N q_\sigma} p_2 H_{p_2} \big[\{m_j^{\sigma}\}\big]
       \nonumber\\
       &&-\frac{ p_1(p_1-1)\beta^2 J_{p_1}^2}{2}(1-q_\sigma)q_\sigma^{p_1-2}-\frac{ p_2(p_2-1)\beta^2 J_{p_2}^2}{2}(1-q_\sigma)q_\sigma^{p_2-2} +O(\beta^3)\Big]
       \nonumber\\
       &&-\sum_i (m_i^\sigma)^2 \Big[ -\frac{1}{1-q_\sigma} -\frac{\beta}{N q_\sigma} p_1 H_{p_1} \big[\{m_j^{\sigma}\}\big]-\frac{ m_i^\sigma\beta}{N q_\sigma} p_2 H_{p_2} \big[\{m_j^{\sigma}\}\big]
       \nonumber\\
       &&-\frac{ p_1(p_1-1)\beta^2 J_{p_1}^2}{2}(1-q_\sigma)q_\sigma^{p_1-2}-\frac{ p_2(p_2-1)\beta^2 J_{p_2}^2}{2}(1-q_\sigma)q_\sigma^{p_2-2} +O(\beta^3)\Big]\; . \nonumber
\end{eqnarray}
We recall here, for the following calculation, that $\{m_i^\sigma\}$ is a metastable TAP state verifying Eqs.~(\ref{eq: TAP eq}) and (\ref{eq: TAP eq(2)}). Consequently the $1^{\rm th}$ order correction in $\beta$ can be rewritten 
\begin{equation}
    -\frac{ m_i^\sigma\beta}{N q_\sigma} p_1 H_{p_1} \big[\{m_j^{\sigma}\}\big]-\frac{ m_i^\sigma\beta}{N q_\sigma} p_2 H_{p_2} \big[\{m_j^{\sigma}\}\big]=-{\beta}\partial_{m_i^\sigma} H_J \big[\{m_j^\sigma\}\big]
\end{equation}
 and it follows that
\begin{eqnarray}
\label{eq: app_constrains}
        h\sum_i (s_i m_i^\sigma-q_\sigma) 
       &=&\sum_i s_i\Big[ -\frac{ m_i^\sigma}{1-q_\sigma} -{\beta}\partial_{m_i^\sigma} H_{ J} \big[\{m_j^\sigma\}\big]
       \nonumber\\
       &&-\frac{ p_1(p_1-1)\beta^2 J_{p_1}^2}{2}(1-q_\sigma)q_\sigma^{p_1-2}m_i^\sigma-\frac{ p_2(p_2-1)\beta^2 J_{p_2}^2}{2}(1-q_\sigma)q_\sigma^{p_2-2}m_i^\sigma +O(\beta^3)\Big]
       \nonumber\\
       &&-\sum_i m_i^\sigma\Big[ -\frac{ m_i^\sigma}{1-q_\sigma} -{\beta}\partial_{m_i^\sigma} H_J \big[\{m_j^\sigma\}\big]
       \nonumber\\
       &&-\frac{ p_1(p_1-1)\beta^2 J_{p_1}^2}{2}(1-q_\sigma)q_\sigma^{p_1-2}m_i^\sigma-\frac{ p_2(p_2-1)\beta^2 J_{p_2}^2}{2}(1-q_\sigma)q_\sigma^{p_2-2}m_i^\sigma +O(\beta^3)\Big]
       \nonumber\\
       &=&\sum_i h_i (s_i-m_i^\sigma)+O(\beta^3)\sum_i(s_i-m_i^\sigma) \;,
\end{eqnarray}
with
\begin{eqnarray}
       &h_i=\frac{-m_i^\sigma}{1-q_\sigma}-\beta \partial_{m_i^\sigma} H_J \big[\{m_j^\sigma\}\big]-\frac{p_1(p_1-1)\beta^2 J_{p_1}^2}{2}(1-q_\sigma)q_\sigma^{p_1-2}m_i^\sigma-\frac{p_2(p_2-1)\beta^2 J_{p_2}^2}{2}(1-q_\sigma)q_\sigma^{p_2-2}m_i^\sigma \; .
\end{eqnarray}
We recover here up to $2^{\rm nd}$ order in $\beta$ not only the TAP free energy but also its constraints on the spherical norm and the local magnetisations with the fields $\lambda$ and $\{h_i\}$:
\begin{eqnarray}
\label{eq: appD3}
  -\beta F_{ J}[\beta,q_o=q_\sigma,\{m_i^\sigma\}]&=&-\beta F_{\rm TAP}\big[\beta,\{ m_i^{\sigma}\}\big]+O(\beta^3)\\
  &=&\ln \Bigg\{  {\rm Tr}_{\{s_i\}}\Big[e^{-\beta H_J[\{s_i\}]-\frac{\lambda}{2}\sum_i (s_i^2-l)-h\sum_i(s_i m_i^\sigma-q_\sigma)}\Big]  \Bigg\} 
  \nonumber\\
  &=&\ln \Bigg\{  {\rm Tr}_{\{s_i\}}\Big[e^{-\beta H_J[\{s_i\}]-\frac{\lambda}{2}\sum_i (s_i^2-l)-\sum_i h_i(s_i-m_i)+O(\beta^3)\sum_i(s_i-m_i^\sigma)+O(\beta^3)\sum_i(s_i^2-1)}\Big]  \Bigg\}\; . \nonumber
\end{eqnarray}

Finally, it is important to point out that the procedure for the high temperature expansion \cite{Biroli1999} is recursive, in other words knowing the fields $\lambda$ and $\{h_i\}$ up to their $(N-1)^{th}$ order in $\beta$ one can derive the $N^{th}$ order correction in $\beta$ of the TAP free energy. Thus, as $-\beta F_{ J}[\beta,q_o=q_\sigma,\{m_i^\sigma\}]$ and  $-\beta F_{\rm TAP}[\beta,\{ m_i^{\sigma}\}]$ share the same fields $\lambda$ and $\{h_i\}$ up to $2^{nd}$  order in $\beta$, their $3^{rd}$ order corrections (and recursively all higher order in $\beta$) are exactly the same. These contributions are sub-extensive in the usual TAP calculation and are not taken into account in our context. We can then write the equality
\begin{eqnarray}
        -\beta F_{ J}[\beta,q_o=q_\sigma,\{m_i^{\sigma}\}] =-\beta F_{\rm TAP}\big[\beta,\{ m_i^{\sigma}\}\big].
\end{eqnarray}

This property is true only under the two assumptions we presented: the reference $\{v_i=m_i^\sigma\}$ is a metastable TAP state at $\beta'$ and the overlap $q_o$ is set to $q_o=q_\sigma$. For example one could set $\Delta H[\{m_i^\sigma\}]_{p_2}=0$ in Eq.~(\ref{eq: appD1}) to derive Eq.~(\ref{eq: appD2}). Yet in this case the previous procedure does not hold, in other words we cannot rewrite the Lagrange multiplier term $h\sum_i (s_i m_i^\sigma-q_\sigma)$ conveniently to derive the constrained free energy under the form in Eq.~(\ref{eq: appD3}). Therefore, in this case,  $O(\beta^3)$ extensive terms still contribute a priori to the constrained free energy.

\bibliographystyle{phaip}

\end{document}